\documentclass[reprint,prl,aps,superscriptaddress,floatfix]{revtex4-2}
\usepackage{times}
\usepackage{graphicx}
\usepackage{amsmath, braket, amsfonts}
\usepackage{amssymb}
\usepackage{sidecap}
\usepackage{bm, color, ulem}
\usepackage{ragged2e}
\usepackage{wrapfig,lipsum,booktabs}
\usepackage[dvipsnames]{xcolor}
\usepackage{pifont}
\newcommand{\cmark}{\ding{51}}
\newcommand{\xmark}{\ding{55}}

\usepackage{siunitx}

\newcommand{\be}{\begin{equation}}
\newcommand{\ee}{\end{equation}}

\newcommand{\figurelabel}[1]{{\color{red}}}

\DeclareUnicodeCharacter{03BD}{{$\nu$}}
\UseRawInputEncoding

\makeatletter
\def\maketitle{
\@author@finish
\title@column\titleblock@produce
\suppressfloats[t]}
\makeatother

\begin{document}
\title{Intervalley coherence and intrinsic spin-orbit coupling in rhombohedral trilayer graphene}
\author{Trevor Arp}
\email{These authors contributed equally: Trevor Arp and Owen Sheekey}
\author{Owen Sheekey}
\email{These authors contributed equally: Trevor Arp and Owen Sheekey} 
\author{Haoxin Zhou}
\email{Current address: Department of Electrical Engineering and Computer Sciences, University of California, Berkeley, California 94720, USA.} 
\author{C.L. Tschirhart}
\email{Current address: Laboratory of Atomic and Solid-State Physics, Cornell University, Ithaca, NY, USA} 
\author{Caitlin L. Patterson}
\author{H. M. Yoo}
\author{Ludwig Holleis}
\author{Evgeny Redekop}
\author{Grigory Babikyan}
\author{Tian Xie}
\affiliation{Department of Physics, University of California at Santa Barbara, Santa Barbara CA 93106, USA}
\author{Jiewen Xiao}
\author{Yaar Vituri}
\author{Tobias Holder}
\affiliation{Department of Condensed Matter Physics, Weizmann Institute of Science, Rehovot, Israel}
\author{Takashi Taniguchi}
\affiliation{International Center for Materials Nanoarchitectonics, National Institute for Materials Science,  1-1 Namiki, Tsukuba 305-0044, Japan}
 \author{Kenji Watanabe}
 \affiliation{Research Center for Functional Materials,
 National Institute for Materials Science, 1-1 Namiki, Tsukuba 305-0044, Japan}
\author{Martin E. Huber}
\affiliation{Departments of Physics and Electrical Engineering, University of Colorado Denver, Denver, CO 80217, USA.}
\author{Erez Berg}
\affiliation{Department of Condensed Matter Physics, Weizmann Institute of Science, Rehovot, Israel}
\author{Andrea F. Young}
\email{andrea@physics.ucsb.edu}
\affiliation{Department of Physics, University of California at Santa Barbara, Santa Barbara CA 93106, USA}
\date{\today}

\begin{abstract}
\end{abstract}
 
\maketitle

\textbf{
Rhombohedral graphene multilayers provide a clean and highly reproducible platform to explore the emergence of superconductivity\cite{zhou_superconductivity_2021,zhou_isospin_2022,zhang_enhanced_2023,holleis_ising_2023} and magnetism\cite{shi_electronic_2020,zhou_half-_2021,zhou_isospin_2022,de_la_barrera_cascade_2022,seiler_quantum_2022,lin_spontaneous_2023,seiler_interaction-driven_2023,han_orbital_2023} in a strongly interacting electron system. 
Here, we use electronic compressibility and local magnetometry to explore the phase diagram of this material class in unprecedented detail.  We focus on rhombohedral trilayer in the quarter metal regime, where the electronic ground state is characterized by the occupation of a single spin and valley isospin flavor. Our measurements reveal a subtle competition between valley imbalanced (VI) orbital ferromagnets and intervalley coherent (IVC) states in which electron wave functions in the two momentum space valleys develop a macroscopically coherent relative phase.  
Contrasting the in-plane spin susceptibility of the IVC and VI phases reveals the influence of graphene's intrinsic spin-orbit coupling\cite{kane_quantum_2005, min_intrinsic_2006, yao_spin-orbit_2007, konschuh_theory_2012}, which drives the emergence of a distinct correlated phase with hybrid VI and IVC character. 
Spin-orbit also suppresses the in-plane magnetic susceptibility of the VI phase, which allows us to extract the spin-orbit coupling strength of $\lambda\approx 50 \mu eV$ for our hexagonal boron nitride-encapsulated graphene system.  
We discuss the implications of finite spin-orbit coupling on the spin-triplet superconductors observed in both rhombohedral and twisted graphene multilayers. 
}

%
%

A defining feature of correlated low-energy electron physics in both moir\'e and crystalline graphene is the large number of closely competing ground states.  
This near-degeneracy is traceable to the approximate SU(4) symmetry within the combined spin and valley `isospin' space, which allows for a large number of broken symmetry phases with nearly degenerate energies.
In experiment, this degeneracy may be lifted either spontaneously or by weak symmetry breaking terms in the Hamiltonian, which compete to determine the ground state.  
Some symmetry breaking effects arise at the single particle level, such as atomic scale spin-orbit coupling\cite{kane_quantum_2005,min_intrinsic_2006,yao_spin-orbit_2007}--though this effect has been presumed to be negligible in the theoretical literature. 
Other symmetries are broken by the inter-particle interactions themselves.  For example, differences between inter- and intravalley scattering reduces the symmetry of the Hamiltonian to $SU(2)_\text{spin}\times U(1)_{K}\times U(1)_{K'}$ representing spin rotation and the independent conservation of charge in the K and K' valleys, respectively\cite{cvetkovic_electronic_2012}. 
Finally, interactions may drive the \textit{spontaneous} breaking of the symmetries that do remain: for example through the formation of intervalley coherent (IVC) ground states characterized by a macroscopically coherent phase between the wave function in the two inequivalent valleys. 
At the theoretical level, the strength of the numerous microscopic parameters that govern breaking of the symmetries cannot be accurately determined from first principles.  As a result, experimental determination of the ground state provides the primary method to constrain the microscopic Hamiltonian. This has proven particularly challenging in twisted graphene multilayers, where experimental irreproducibility has hampered efforts to reliably determine the phase diagram. 

Rhombohedral graphene multilayers provide a structurally simple and experimentally reproducible platform, enabling a concrete connection between precision measurements and many-body theory.
In rhombohedral graphene alone, experiments have revealed signatures of symmetry breaking states that include nematics\cite{mayorov_interaction-driven_2011,lin_spontaneous_2023,holleis_ising_2023,zhang_enhanced_2023}, spin and orbital magnets\cite{shi_electronic_2020, zhou_half-_2021, zhou_isospin_2022,
de_la_barrera_cascade_2022, seiler_quantum_2022, winterer_ferroelectric_2023, han_orbital_2023, han_correlated_2023}, 
and superconductors\cite{zhou_superconductivity_2021, zhou_isospin_2022, zhang_enhanced_2023, holleis_ising_2023}. 
However, much of the observed phenomenology---particularly the role of spin in the superconducting phases---remains unexplained. 
A particularly striking puzzle is the stability of superconductivity in hexagonal boron nitride encapsulated Bernal bilayer graphene, which shows no superconductivity at zero magnetic field but in which a spin-polarized superconducting state emerges above a threshold magnetic field applied in the plane of the sample\cite{zhou_isospin_2022}.
Spin-valley locked superconductivity is also induced (at zero magnetic field) by supporting the bilayer on a WSe$_2$\cite{zhang_enhanced_2023,holleis_ising_2023} substrate, which is known to induce a large Ising-type spin-orbit coupling\cite{island_spinorbit-driven_2019}---further pointing to the role of spin in the superconducting phase diagram.  
Of course, in the absence of a clear understanding of what stabilizes the superconducting states, consensus has also proved elusive regarding the underlying mechanism, with both all-electronic\cite{you_kohn-luttinger_2022,ghazaryan_unconventional_2021,ghazaryan_multilayer_2023,cea_superconductivity_2022,wagner_superconductivity_2023,chatterjee_inter-valley_2022,dong_superconductivity_2021,dong_signatures_2023,dong_spin-triplet_2022} and phonon-mediated\cite{chou_acoustic-phonon-mediated_2021} mechanisms remaining viable in light of current experimental data.

\begin{figure*}[ht!]
    \centering
    \includegraphics{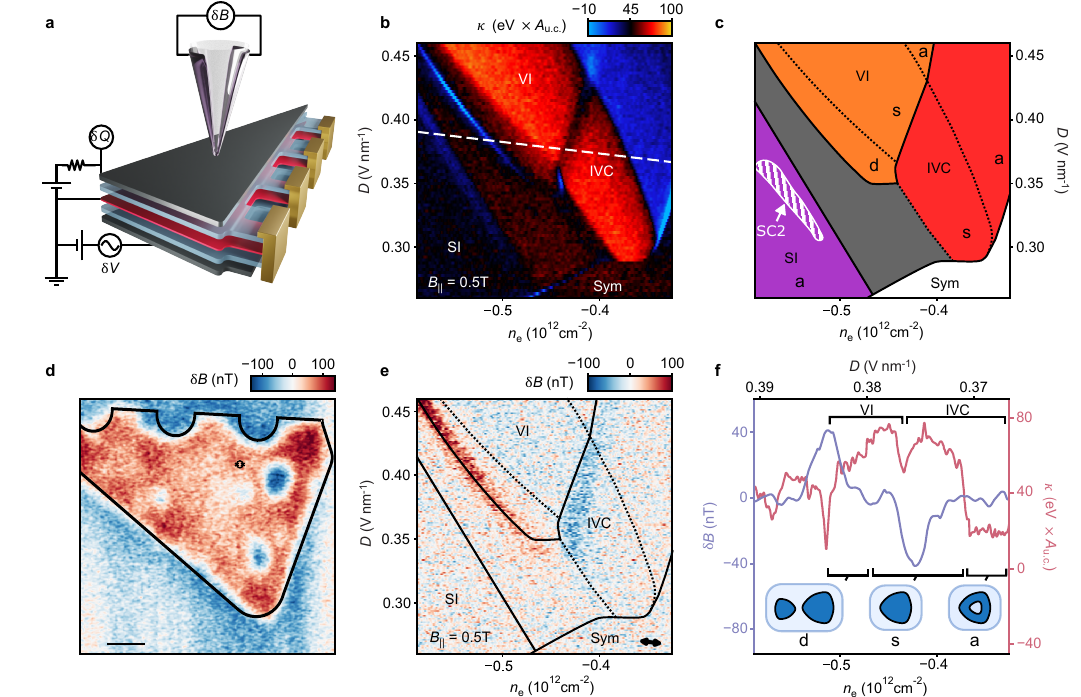}
    \caption{
    \textbf{Thermodynamics of  rhombohedral trilayer graphene in the hole-doped quarter metal regime.}
    \textbf{(A)} Schematic of the measurement geometry. The bottom gate is modulated at kHz frequencies with amplitude $\delta \mathrm{V}$.  A transistor amplifier connected to the top gate measures the modulated charge $\delta Q$, from which the inverse compressibility, $\kappa$, is derived\cite{eisenstein_negative_1992}. A scanning SQUID magnetometer measures the local magnetic field $\delta B$. 
    (\textbf{B}) $\kappa$, as a function of carrier density $n_e$ and applied displacement field $D$ at $T \approx 20$~mK and $B_{\parallel} = 0.5$~T. 
    (\textbf{C})  Schematic phase diagram showing competing phases, classified by isospin polarization and Fermi surface topology.   Uppercase labels denote the isospin polarization, which can be valley imbalanced (VI),  intervalley coherent (IVC), spin imbalanced (SI), or isospin symmetric (Sym); lower-case labels denote simple (s), annular (a), or disjoint (d) Fermi surface topology.    
    The region marks SC2 denotes the superconductor reported previously\cite{zhou_superconductivity_2021} near the edge of the SI phase.  As discussed in the text, the isospin order of the gray region cannot be determined conclusively.
    (\textbf{D}) Spatial image of $\delta B$ taken $\sim 300$~nm above the rhombohedral trilayer. The image is acquired at $B_\perp=15$~mT, $n_{e} = -0.67 \times 10^{12}$~cm$^{-2}$ and $D=0.55$~V/nm, corresponding to a first order transition into an orbitally magnetized phase.  The black outline indicates the design geometry of the dual gated region of the device. Scale bar is 1~$\mu$m.
    (\textbf{E}) $\delta B$ at $B_{\parallel} = 0.5$T and $T = 300$~mK, acquired at the position marked in panel \textbf{C}. 
    Lines indicate the boundaries between phases from \textbf{A}, with solid lines indicating negative compressibility peaks and dashed lines compressibility steps associated with first order and Lifshitz transitions, respectively.  
    The arrow at bottom right shows the direction of the $\delta \mathrm{V}$ modulation.  
    (\textbf{F}) $\delta B$ (blue) and $\kappa$ (red) along the dashed line shown in panel A. The regions with VI and IVC order are separated by a first order phase transition (visible in $\kappa$), as well as a change in magnetization indicated by a peak in $\delta B$.  Changes in Fermi surface topology occur at Lifshitz transitions ---steps in $\kappa$--and are not accompanied by features in $\delta B$.  
    \figurelabel{fig:1}
    }
    \label{fig:1}
\end{figure*}

Here we combine global charge sensing and local magnetometry measurements in rhombohedral trilayer graphene to explore the nature of the isospin ferromagnetic phases that emerge as the material is doped through the Van Hove singularities bracketing the neutrality point. 
Fig.~\ref{fig:1}A shows a schematic of our measurement geometry. 
Our rhombohedral trilayer graphene sample is encapsulated between hexagonal boron nitride gate dielectrics and graphite flakes which serve as electrostatic gates to control the total charge carrier density  $n_e$ and applied displacement field $D$ (details of sample fabrication, and transport measurements from the same device, were reported previously\cite{zhou_half-_2021,zhou_superconductivity_2021}).  
We use a cryogenic transistor amplifier\cite{ashoori_single-electron_1992} connected to the top gate to monitor the charge $\delta Q$ induced by the  modulated bottom gate voltage $\delta \mathrm{V}$. 
$\delta Q$  is proportional to the spatially averaged inverse compressibility $\kappa = \partial \mu / \partial n_{e}$. 
In addition, we use a scanning superconducting quantum interference device to image the local magnetic field, $\delta B$, that arises in response to the same modulated voltage. For purely out of plane magnetic moments, $\delta B$ is proportional to the gate-modulated change in magnetization. 

\section{Intervalley coherent quarter metal}\label{sec:IVC}

Figure~\ref{fig:1}B shows $\kappa$ measured for hole doping, with $B_\parallel=0.5$~T  magnetic field applied to ensure in-plane spin polarization. As shown previously, in the low-$|n_e|$, low $D$ extreme of this range the electron system is in a symmetric (`Sym') phase that preserves isospin symmetry.  
At low $D$ and high-$|n_e|$, the system is a valley unpolarized, spin-polarized half-metal\cite{zhou_superconductivity_2021} in which the two Fermi seas have annular topology.  We denote this phase SI in the schematic in Fig.~\ref{fig:1}C, as it is an example of a spin-imbalanced phase.  Spin-triplet superconductivity was reported at the low-$|n_e|$ extreme of this `half-metal' SI phase\cite{zhou_superconductivity_2021}. 
In the rest of the phase space spanned by Figs.~\ref{fig:1}B,C measurements of quantum oscillations have shown\cite{zhou_superconductivity_2021} evidence of valley polarization.  In the regions rendered in red and blue in Fig. \ref{fig:1}B, quantum oscillations show a single Fermi sea with either simple or annular topology---a quarter metal. Quantum oscillations in the intermediate phase between the SI and quarter metal phases were interpreted as indicating partial valley polarization\cite{zhou_half-_2021,zhou_superconductivity_2021}.

In the present work, we improve upon previous compressibility measurements through the use of a thermal decoupling capacitor between sample and cryogenic amplifier and an additional cryogenic amplification stage at 4K (see Methods).  The data in Fig.~\ref{fig:1}B reveals both `step' features, associated with Lifshitz transitions, and  negative compressibility spikes, associated with first order phase transitions, that were not previously reported.  Most strikingly, a dip in 
$\kappa$ bisects both the simple and annular quarter metal regime. As indicated in Fig.~\ref{fig:1}C, we ascribe this feature to a transition between distinct valley imbalanced (VI) and inter-valley coherent (IVC) quarter metal phases.  

The contrast between these phases can be visualized directly using local magnetometry. Fig.~\ref{fig:1}D shows a spatial map of $\delta B$ acquired $300$~nm above the sample layer at a point in the $n_e, D, B_\perp$ and $B_\parallel$-tuned parameter space where the system is at a transition between phases with contrasting valley population imbalance. The $\delta B$ image of Fig.~\ref{fig:1}D thus represents the change in fringe magnetic field associated with a change in the sample's orbital magnetic moment.
We find the spatial structure of $\delta B$ to be largely independent of $n_e$ and $D$, indicating that the phase transitions observed in our global $\kappa$ measurements occur in the entire sample simultaneously. An exception to this rule is the behavior of transitions associated with spin imbalance changes as a function of $B_\parallel$ (Extended Data Fig.~\ref{fig:ext_data_half_metal}); specifically, we observe the signatures of in-plane spin moments at these transitions when $|B_\parallel|>|B_\perp|$.

Fig.~\ref{fig:1}E shows $\delta B$ measured at a single spatial position as a function of $n_e$ and $D$ for $B_\parallel = 0.5$~T.  The position is chosen to minimize sensitivity to the in-plane spin magnetic moment. The spatial uniformity of the sample then allows us to interpret $\delta B$ as equivalent to the modulated out-of-plane magnetic moment, $\delta m_z$, up to a multiplicative constant.
We assume the spins are fully polarized in the plane, so that $\delta m_z$ arises only from the intrinsically out-of-plane orbital magnetization.\cite{das_unconventional_2023} 

Comparing $\delta B$ (Figs.~\ref{fig:1}E) and $\kappa$ (Figs.~\ref{fig:1}B) reveals the nature of the two quarter metal phases.  Only transitions where the valley imbalance---and thus orbital moment---changes show contrast in $\delta B$. 
Within the quarter metal phase, the region to the left of the first order transition shows the signatures of a finite orbital moment, and we assign it to a valley imbalanced phase 
characterized by the complete- or near-complete polarization of the electron system into a single valley.  
Additional transitions are visible in $\kappa$ within the VI phase, which we assign to Lifshitz transitions in the fermi sea topology that are not accompanied by any change in isospin symmetry.  
Fermi sea topology may be simple ($s$), annular ($a$), or disjoint ($d$)--where $d$ means there exist fermi pockets of different sizes but the same carrier sign.  
These transitions manifest as steps in the electronic compressibility without signature in $\delta B$, as shown in Fig.~\ref{fig:1}F. 

Outside of this VI phase, none of the phases in the range of Figs.~\ref{fig:1}B,D have finite orbital moment.  This is expected for the Sym and SI phases, but is novel in the case of the simple- and annular quarter metals, as well as the region, indicated in gray in Fig.~\ref{fig:1}C separating the quarter metal and SI phases.   
In the case of quarter metals only an intervalley coherent (IVC) state, characterized by a coherent superposition of states in the two valleys, is consistent with a vanishing orbital moment and full flavor polarization.  We do not make a definitive assignment for the phase marked in gray.  This phase is spin polarized and valley balanced (Extended Data Fig.~\ref{fig:ext_SC2}), but it disappears from the phase diagram by $B_\perp \approx 100$mT preventing detailed study of quantum oscillations. 
While Hartree Fock calculations (Extended Data Fig.~\ref{fig:newHF}) show that an intervalley coherent state may be energetically favorable in this regime, the close energetic competition between the IVC state and other spin-polarized states that break orbital symmetries but preserve conservation of charge in the two valleys--for example, nematics--cannot be ruled out.

\begin{figure}[ht]
    \centering
    \includegraphics[width = \linewidth]{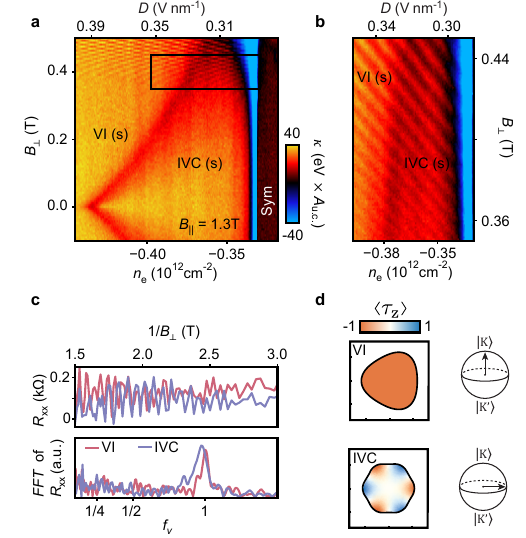}
    \caption{\textbf{Intervalley coherent quarter metal.}
    (\textbf{A}) $\kappa$ as a function of $n_{e}$ and $B_{\perp}$ along a trajectory that traverses the VI, IVC and Sym phases at $B_{\parallel} = 1.3$T and $T = 20$~mK. 
    (\textbf{B}) Higher resolution measurement over the boxed region of panel A. 
    (\textbf{C}) $R_\text{xx}(1/B_{\perp})$ (top) and its Fourier transform (bottom), plotted as a function of the normalized frequency $f_\nu$. Curves measured in the VI and IVC phases both show a single peak at $f_\nu=1$ indicating a single flavor polarized Fermi surface. 
    (\textbf{D}) Schematic depiction of the valley polarization of the  IVC and VI quarter metal phases (left) and their representation on a valley Block sphere (right).  For the IVC phase, (net) in-plane valley polarization, e.g. $\langle \tau_x\rangle = 1$,  corresponds to a momentum dependent valley polarization $\langle \tau_z\rangle$. 
    \figurelabel{fig:2}}
    \label{fig:2}
\end{figure}

To substantiate the assignment of IVC order in Fig.~\ref{fig:1}E, we analyze $\kappa$, measured across the phase boundary separating VI and IVC phases with simple Fermi surface topology, as a function of magnetic field .  The first order nature of the transition implies a Clausius-Clapeyron-type relation between the critical density $n_e^*$ and the magnetic moment difference $\Delta \vec m$ between the two phases,   
\begin{equation}
     \frac{\Delta \vec m}{\Delta \mu}= \frac{dn_e^{\ast}}{d\vec B}, 
\label{eqn:clausiusclapeyron}
\end{equation}
where $\Delta \mu$ is magnitude of the chemical potential 
jump between the two phases and $\vec m\parallel \vec B$.  
Fig.~\ref{fig:2}A shows the IVC to VI transition as a function of $B_\perp$. To isolate the effects of orbital magnetism, the measurements are again performed in a large, fixed in-plane magnetic field $B_\parallel=1.3T\gg B_\perp$. 
The  boundary between the VI and IVC phase evolves rapidly with $B_\perp$ in favor of the VI phase, consistent with $m_\perp^\text{VI}>m_\perp^\text{IVC}$.  The  sharp cusp in the critical density at $B_\perp = 0$ implies a divergent orbital magnetic susceptibility in the VI phase.   

\begin{figure*}[ht]
    \centering
    \includegraphics{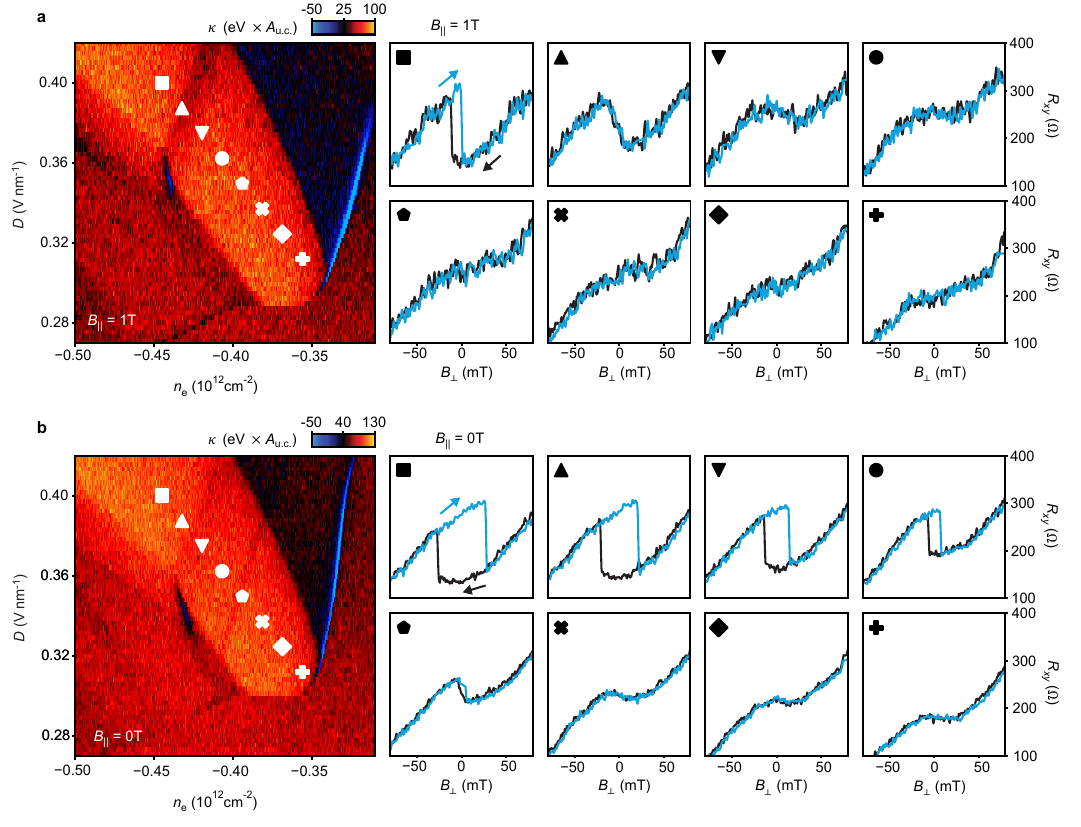}
    \caption{\textbf{Anomalous Hall effect.}
    (\textbf{A}) Resistance measured in an R$_{\text{XY}}$-like configuration at different points in the quarter metal phase for $T \approx 20$~mK and B$_{\parallel} = 1$~T. Anomalous Hall effect is observed only in the VI phase. 
    (\textbf{B}) Resistance in the quarter metal phase at $T \approx 20$~mK and B$_{\parallel} = 0$~T.  Both VI and $IVC_z$ phases show anomalous Hall signal. 
     \figurelabel{fig:ext-hysteresis}
    }
    \label{fig:ext-hysteresis}
\end{figure*}

In contrast, the IVC-s/Sym phase boundary remains at fixed density as $B_\perp$ is tuned, consistent with $m_\perp^\text{IVC}=m_\perp^\text{Sym}=0$.  
Notably, quantum oscillations show no break or phase shift across the IVC to VI transition (Fig.~\ref{fig:2}B), implying consistent Fermi surface topology between the two phases.  
This is confirmed in Fig.~\ref{fig:2}C, which shows $R_\text{xx}$ traces within the two phases. Data is shown both as a function of $1/B_\perp$ and of $f_\nu$, the frequency normalized to the total Luttinger volume. Both phases show a single peak at $f_{\nu} = 1$, corresponding to a single Fermi surface, a behavior which extends across the entire simple quarter metal regime (see Extended Data Fig.~\ref{fig:ext_data_QOs}). 
Finally, Hall measurements show hysteretic anomalous Hall effect only within the VI phase but not in the IVC phase (Extended Data Fig.~\ref{fig:ext-hysteresis}).
Taken together, these data provide strong evidence for intervalley coherent phases in which the ground state is an equal superposition of wavefunctions in the two valleys (shown in Fig.~\ref{fig:2}D), leading to a cancellation of the net orbital moment.  We note that in real space, the IVC phases described here are expected to show lattice-scale charge density wave order, making them detectable by the same scanning tunneling microscopy techniques that have identified intervalley coherent order in monolayer graphene at high magnetic fields\cite{coissard_imaging_2022,liu_visualizing_2022} and moir\'e graphene systems\cite{nuckolls_quantum_2023,kim_imaging_2023}. 

\section{Spin-orbit Coupling} \label{sec:SOC}

 \begin{figure*}[ht!]
    \centering
    \includegraphics{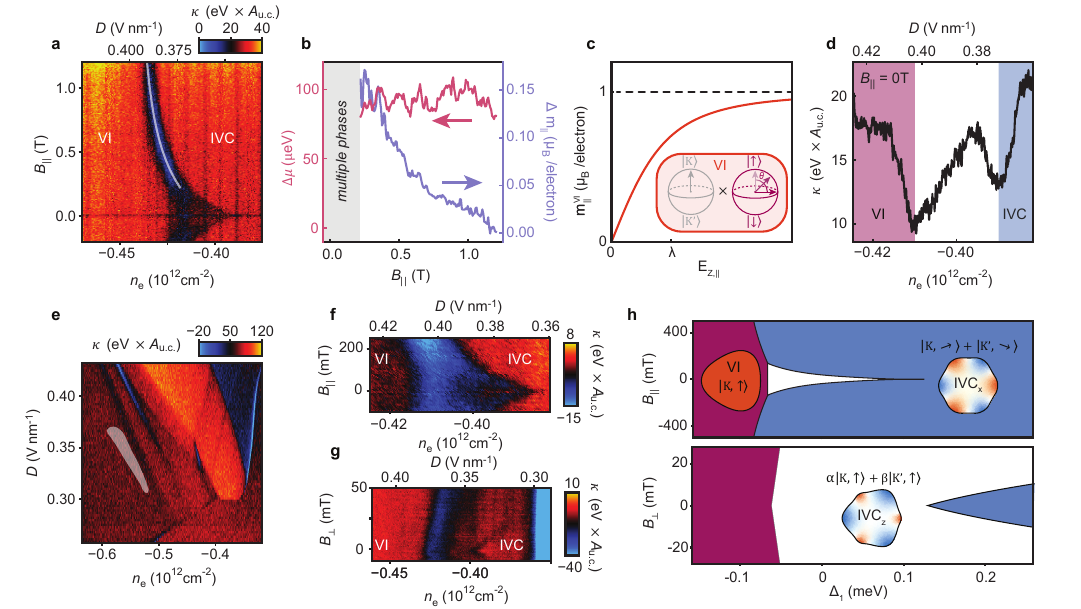}
    \caption{\textbf{Effects of spin-orbit coupling.}
    (\textbf{A}) $\kappa$ across the VI to IVC transition as a function of $B_{\parallel}$ at $T\approx 300$mK. The white line shows the fit used to extract $\lambda\approx$~46~$\mu$eV described in the main text. 
    (\textbf{B}) $\Delta \mu$ as determined by integrating the dip in $\kappa$ (red) and $\Delta m_\parallel=\Delta \mu (\partial n_e/\partial B_\parallel)$ (blue).  
    (\textbf{C}) In-plane spin magnetic moment of the VI phase in the presence of spin-orbit coupling as a function of $B_\parallel$. Inset shows the evolution of the VI phase in the Bloch sphere representation; the spin cants as a function of $B_\parallel$ while the valley remains polarized.
    (\textbf{D}) $\kappa$ across the VI to IVC transition at B=0, showing the intermediate phase. 
    (\textbf{E}) $\kappa$ in the hole-doped quarter metal regime at $B_\parallel=B_\perp=0$. 
    White overlay denotes range of SC2.\cite{zhou_superconductivity_2021} 
    (\textbf{F}) $\kappa$ across the VI to IVC transition as a function of $B_{\parallel}$ showing the suppression of the intermediate phase at low in-plane field. 
    (\textbf{G}) $\kappa$ as a function of $B_{\perp}$.  The intermediate phase rapidly grows with $B_\perp$, displacing the IVC phase completely by 50 mT. 
    (\textbf{H}) Phase diagram of a phenomenological model, described in the supplementary information, that accounts for the competing effects of magnetic field, exchange energy, Hund's coupling, and spin-orbit effects.  
    Insets show Fermi surfaces, color coded by $\langle \tau_z\rangle$ as in Fig. \ref{fig:2}D.  
    \figurelabel{fig:3}
    }
    \label{fig:3}
\end{figure*}

\begin{figure}[ht]
    \centering      \includegraphics{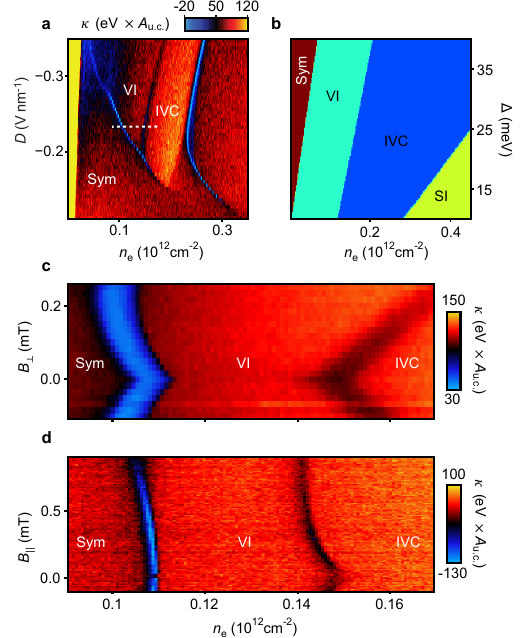}
    \caption{\textbf{Intervalley coherence for electron doping}
    (\textbf{A}) $\kappa$ as a function of carrier density $n_e$ and applied displacement field $D$ for electron doping  at $T \approx 20$~mK and $B_{\parallel} = 0$. 
    (\textbf{B}) Hartree-Fock phase diagram, showing the ground state isospin polarization as a function of $n_{e}$ and the interlayer potential $\Delta$, which is approximately proportional to $D$\cite{zhang_direct_2009}.
    (\textbf{C}) Evolution of the Sym, VI, and IVC phases along the dashed line in panel A as a function of $B_\perp$ and 
    (\textbf{D}) $B_{\parallel}$. 
    \figurelabel{fig:4}}
    \label{fig:4}
\end{figure}

The phase diagrams presented above are all acquired at large in-plane magnetic field. However, many of the phase boundaries show  non-trivial behavior as $B_\parallel$ is reduced. 
Fig.~\ref{fig:3}A shows the VI to IVC boundary studied in  Figs.~\ref{fig:1} and \ref{fig:2} as a function of $B_\parallel$.
At high $B_\parallel$, the transition is nearly $B_\parallel$ independent, but at low $B_\parallel$ the IVC phase is increasingly disfavored relative to the VI phase. As shown in Fig.~\ref{fig:3}B, $\Delta \mu$ is independent of $B_\parallel$ over much of the range.  It follows from Eq. \eqref{eqn:clausiusclapeyron} that the evolution of $n_e^\ast$ implies a finite, and $B_\parallel$ dependent, difference in $m_\parallel$ between the two phases. Given the near complete layer polarization of the electron wave function at the high $D$ of the experiment, $m_\parallel$ arises exclusively from the electron spin.  
The finite $\Delta m_\parallel$ is thus unexpected, as both the VI-s and IVC-s phases are in principle fully spin polarized, and are expected, in the absence of spin-orbit coupling, to have identical and divergent susceptibility to spin polarization in any direction. 

Intrinsic spin-orbit coupling\cite{kane_quantum_2005}---typically neglected in theoretical treatments of graphene many body physics---provides a natural explanation for the anomalous behavior of $\Delta m_\parallel$. 
Projected into the low-energy bands near the K-points of the Brillouin zone, spin-orbit coupling in graphene takes the form\cite{kane_quantum_2005} $H_\text{KM}=\frac{\lambda}{2} \sigma_z\tau_z s_z$ (where $\sigma_z$, $\tau_z$, and $s_z$ are Pauli matrices in the sublattice, valley, and spin space respectively).  
Theoretical estimates of $\lambda$ range over two orders of magnitude, from 1 to 100~$\mu$eV 
\cite{kane_quantum_2005,min_intrinsic_2006,konschuh_theory_2012,yao_spin-orbit_2007,ochoa_spin-orbit_2012}.  
Recent experiments in mono- and bi-layer graphene, however, suggest that $\lambda$ falls somewhere in the range 40 to 80~$\mu$eV \cite{sichau_resonance_2019, banszerus_observation_2020,kurzmann_kondo_2021}.   

Spin-orbit coupling has contrasting effects on the IVC and VI phases.  At high displacement field in the layer (and therefore sublattice) polarized limit, Kane-Mele spin-orbit coupling reduces to a staggered effective Zeeman energy  which is opposite in sign in the two valleys (the `Ising' spin-orbit coupling familiar from other honeycomb systems with broken inversion symmetry).  
For finite $\lambda$, a spin-polarized, valley-imbalanced phase may lower its energy by aligning its spin in the spin-orbit favored, out-of-plane direction.  As $B_\parallel$ increases, the spins in the VI phase cant into the plane.  Taking 
$E_{z,\parallel}=g \mu_B B_\parallel$, 
the in-plane component of the canted spin is given by $\text{m}_\parallel^\mathrm{VI} = E_{z,\parallel}/ [E_{z,\parallel}^2 + \lambda^2 ]^{1/2}$, plotted in Fig.~\ref{fig:3}C.   
In contrast, the energy of the valley balanced IVC phase is independent of  spin-orbit coupling to first order in $\lambda$.   
The IVC phase may thus polarize in-plane for arbitrarily small $B_\parallel$, with a divergent susceptibility and subsequent energy gain that is linear in the applied field. 
A fit to the measured $n_e^\ast$, shown in overlay on Fig. \ref{fig:3}A, gives $\lambda \approx 46\mu$eV, in agreement with the experimental literature where this coupling was measured in different contexts and using different experimental techniques\cite{sichau_resonance_2019, banszerus_observation_2020, kurzmann_kondo_2021}. 

\begin{figure*}[ht]
    \centering
    \includegraphics{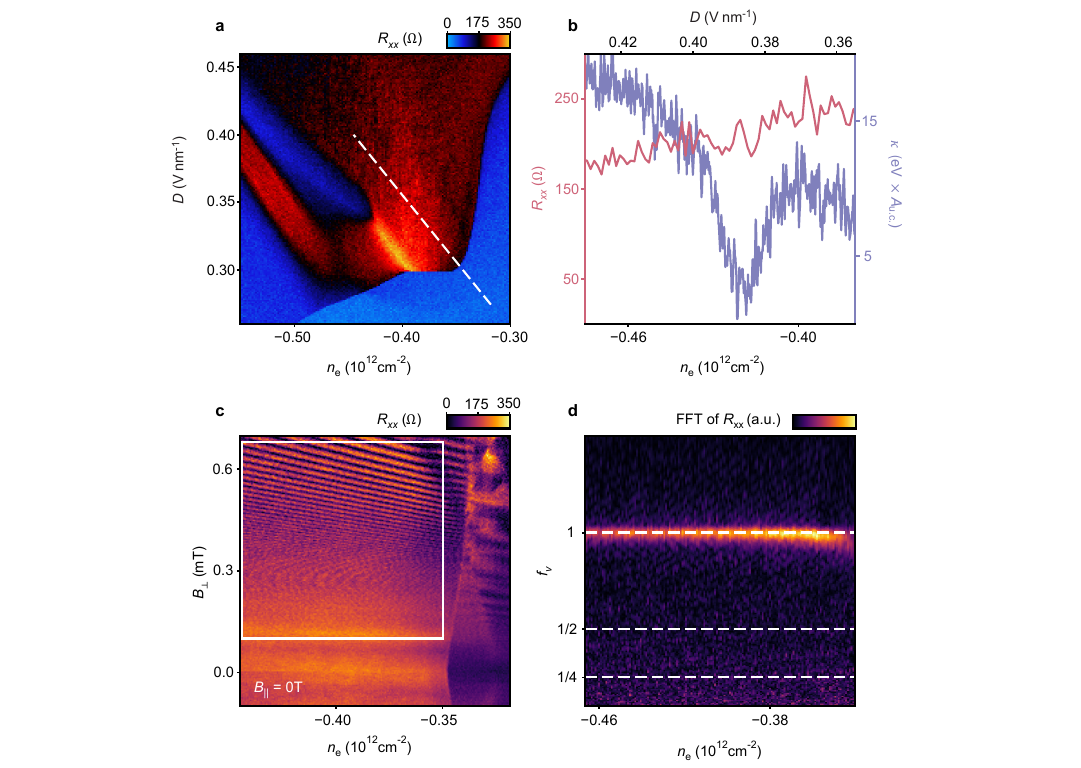}
    \caption{\textbf{Longitudinal Resistance of Quarter Metals.}
    \textbf{(A)} $R_{xx}$ for hole doping at $B_{\parallel} = 0$T at T=20mK. The IVC and VI phases are not distinct in longitudinal resistivity.  
    (\textbf{B}) $R_{xx}$ compared to $\kappa$ as a function of $n_{e}$ and $D$ along the trajectory indicated in \textbf{A}.  The first order phase transitions between VI and IVC do not generate contrast in $R_{xx}$.
    (\textbf{C}) 
    $R_{xx}$ along the trajector in panel A, as a function of $B_{\perp}$ with $B_\parallel=0$.  As explained in the main text, this range largely spans the VI and $IVC_z$ phases. 
    (\textbf{D}) FFT of the data outlined by the white box in B, with the FFT taken as a function of $1/B_{\perp}$ at each value of $n_{e}$.  $f_\nu$ is the FFT frequency normalized to $n_{e}$. 
    \figurelabel{fig:ext\_data\_QOs}
    }
    \label{fig:ext_data_QOs}
\end{figure*}

Near $B_\parallel=0$, spin-orbit coupling produces even richer effects. Measurements across the VI to IVC transition reveal an intermediate phase, separated by compressibility dips from the IVC and VI phases (Figs.~\ref{fig:3}D-E). 
Quantum oscillations in this phase indicate that it is also a quarter metal with a single Fermi surface (Extended Data Fig.~\ref{fig:ext_data_QOs}). As shown in Figs. \ref{fig:3}F-G, the intermediate phase has in- and out-of plane susceptibilities that are intermediate between those of the IVC and VI phases.  This is consistent with a intervalley coherent phase with partial valley  imbalance, endowing it with a finite orbital moment but a smaller suppression of in-plane susceptibility relative to the VI phase. 

In fact, such a phase is also accounted for by the effect of spin-orbit coupling on the spin-polarized IVC quarter metal phase. 
In the absence of spin-orbit coupling, the spin magnetic moment can point in any direction. Spin-orbit coupling induces either an easy-plane or an easy-axis spin anisotropy within any spin polarized phase, including the IVC phases. If the average spin moment is in-plane, the system will lower its energy by slightly canting the spins out of the plane in an opposite direction in the two valleys; we dub this phase IVC$_\mathrm{x}$.
The energy gain associated with the canting scales as $-\lambda^2/J_H$, where $J_H$ is the ferromagnetic inter-valley Hund's coupling. If the spin is pointing out of the plane, on the other hand, the system lowers its energy by introducing a small imbalance between the population of the two valleys. We dub this phase IVC$_\mathrm{z}$. 
The associated energy gain is of the order of $-\chi_v \lambda^2$, where $\chi_v$ is the valley susceptibility of the IVC phase in the absence of spin-orbit coupling. As we approach the IVC to VI phase transition from the IVC side, and assuming that the transition is not too strongly first order, 
$\chi_v$ is expected to increase, and hence the out-of-plane polarized IVC$_\mathrm{z}$ phase is favored over the in-plane polarized IVC$_\mathrm{x}$ phase. 
Fig. \ref{fig:3}H shows the phase diagram of a simple model that captures the relative energetics of the IVC$_\mathrm{x}$, IVC$_\mathrm{z}$, 
and VI phases as a function of $B_\parallel$,  $B_{\perp}$, and $\Delta_{\rm{V}}$, the exchange splitting between VI and IVC phases. 
In the vicinity of $n_e^*$, we take $\Delta_{\rm{V}}\propto n_e-n_e^*$, making $\Delta_{\rm{V}}$ a reasonable proxy for the carrier density tuned in experiment.  This assumption is reproduced by Hartree-Fock calculations which show a large $\chi_V$ in the vicinity of the VI to IVC phase transition (see Extended Data Fig. \ref{fig:HFSOC}).  
Despite its simplicity, the model reproduces all the key features of the experimental data, and quantitatively reproduces the in- and out-of-plane fields required to appreciably tune the balance between phases.

\section{Electron doping}

Theoretically, the competition between VI and IVC states is governed by the interplay of exchange and kinetic energy\cite{chatterjee_inter-valley_2022}; VI states effectively minimize the exchange energy, but the momentum-dependent valley polarization of the IVC phases reduces the kinetic energy in the presence of trigonal warping (see Fig.~\ref{fig:2}D).
For hole doping, where the Van Hove singularity occurs at a finite-momentum saddle point, there is a large kinetic energy contribution from the trigonally warped pockets near the band maximum, accounting for the dominance of IVC order at lower hole density as observed in our experiment. This contrasts with the situation for electron doping, where the bands have a nearly quartic dispersion and the Van Hove singularity is very close to the band edge for a large range of $D$. Within the electron band, the effects of trigonal warping are most pronounced at high density, contrasting with the hole band.

Fig. \ref{fig:4}A shows $\kappa$ measured for electron doping ($n_e>0$). As at hole doping, we observe a first-order phase boundary separating two phases with a single Fermi surface, in addition to the previously reported\cite{zhou_half-_2021} first-order transitions separating the quarter metal region from an isospin symmetric (Sym) phase at low $n_e,D$ and a spin-polarized half metal state (SI) at high $n_e$.  
Fig.~\ref{fig:4}B shows a phase diagram calculated within the Hartree-Fock approximation, plotted as a function of $n_e$ and the interlayer potential difference $\Delta$, which is proportional to $D$ (see Methods and Supplementary information for further details). The simulations predict a valley imbalanced quarter metal at low $n_e$ and a VI phase at intermediate $n_e$, and we assign these labels to the two quarter metal phases in Fig.~\ref{fig:4}A.    

To verify this hypothesis, we track the phase transitions along the dotted line in Fig. \ref{fig:4}A as a function of magnetic field.  The VI phase has divergent susceptibility to $B_\perp$, both as compared to the IVC phase and the Sym phase (Fig. \ref{fig:4}C). 
The behavior of the VI phase is consistent with a vanishing in-plane susceptibility at $B_\parallel=0$,and the gradual canting of the spin moment with increasing $B_\parallel$. This manifests at the VI to IVC transition in Fig. \ref{fig:4}D, where the curvature produces an estimate of $\lambda \approx 32\mu$eV. The VI-Sym boundary is also consistent with this mechanism, following a hyperbolic trajectory from which we extract $\lambda\approx 65\mu$eV (a summary of the $\lambda$ fitting is presented in the methods, and in Extended Data Fig.~\ref{fig:ext_data_electron}). These values of $\lambda$ are consistent with each other and with literature values within the large systematic errors (detailed in the Methods and supplementary information) inherent to this fitting procedure. We note that the VI to IVC transition in the electron-doped quarter metal was missed in the lower resolution measurements of Ref.~\cite{zhou_half-_2021}; as a result, anomalous Hall measurements in the VI phase and the divergent in-plane spin susceptibility of the IVC phase were mistakenly attributed to a single, valley polarized phase with vanishing spin anisotropy. 

\section{Discussion}

Theoretically, intervalley coherent fluctuations may generate attractive electronic interactions even in the absence of acoustic phonons. 
Concrete scenarios for purely electronic superconductivity from this mechanism have been proposed for both rhombohedral graphene multilayers\cite{chatterjee_inter-valley_2022,dong_signatures_2023} and twisted graphene multilayers\cite{lee_theory_2019}. 
In twisted bilayer and trilayer graphene IVC order was recently detected\cite{nuckolls_quantum_2023,kim_imaging_2023} in regions of the parameter space where superconductivity also occurs\cite{cao_unconventional_2018,park_tunable_2021,hao_electric_2021}. 
While this scenario is appealing, it has been in tension with the fact that superconductivity is so widespread in graphene systems: superconductivity is observed in a broad range of twist angles and densities in graphene multilayers\cite{arora_superconductivity_2020}, including far from where IVC order has been detected\cite{zhang_promotion_2022,park_robust_2022}.  

Our observations open up the possibility that intervalley coherence may be just as widespread as superconductivity. Our measurements are able to detect IVC order only in the quarter metal regime where the space of possible phases is already significantly constrained. 
Outside this regime, we cannot conclusively distinguish between intervalley coherence and other theoretically predicted phases, for example, nematic orders with no net orbital magnetization \cite{koh_correlated_2023, zhumagulov_emergent_2023}.
of particular interest, both the phase represented in gray in Fig.~\ref{fig:1}F, adjacent to the spin-triplet SC2, and the spin-unpolarized phase adjacent to the Pauli-limited SC1, are compatible with intervalley coherence given existing experimental constraints.

We also highlight the observation of a large valley susceptibility within the IVC phase, despite the first order nature of the VI to IVC transition. 
Superconductivity is observed on the disordered side of phase transitions to potential IVC states\cite{holleis_ising_2023}. 
Naively, the first order nature of these transitions argues against the importance of fluctuations.  However, our results suggest that in rhombohedral graphene these transitions are sufficiently weakly first order so as to allow for a large susceptibility.  

We finally comment on the implications of finite spin-orbit coupling, which has not been extensively discussed in the context of many-body physics 
in graphene. A number of experimental reports in twisted graphene multilayers have observed anomalously large effects of in-plane field on orbital ferromagnetic order\cite{sharpe_evidence_2021, kuiri_spontaneous_2022}.  This was presumed not to originate from intrinsic spin-orbit coupling due to its negligible magnitude. 
However, our estimate for $\lambda\approx 50 \mu$eV is non-negligible when compared to the energy differences between competing magnetic states---expected to be in the few 100~$\mu$eV range\cite{chatterjee_inter-valley_2022, koh_correlated_2023,huang_spin_2023}---and is considerably \textit{larger} than the superconducting pairing energy inferred from the transition temperature in rhombohedral multilayers\cite{zhou_superconductivity_2021,zhou_isospin_2022,zhang_enhanced_2023,holleis_ising_2023}.  

Spin-orbit coupling is likely to play a particularly important role in the spin-triplet superconducting states observed in a variety of graphene  multilayers\cite{zhou_superconductivity_2021,zhou_isospin_2022,cao_pauli-limit_2021}, selecting the direction of the spin of the Cooper pairs and even forbidding certain phases. 
For example, in rhombohedral bi- and trilayers triplet superconductivity arises from a spin-imbalanced phase, identified from quantum oscillations as an annular half metal in the trilayer and partially spin polarized phase in the bilayer.  
In the presence of spin-orbit coupling, the phase diagram of these spin-imbalanced states is expected to be similar to that of the IVC quarter metal shown in Fig.~\ref{fig:3}H, i.e. with competing easy-axis and easy-plane spin ferromagnetic phases. 
Interestingly, while easy-plane phases maintain valley balance, easy axis phases develop finite valley polarization, which is strongly pair breaking for a superconductor whose order parameter carries zero momentum.  
It is notable that in bilayer graphene, superconductivity emerges at values of $B_\parallel$ where the spins become fully in-plane polarized, at the same field scale where we find a IVC$_\mathrm{z}$ to IVC$_\mathrm{x}$ transition in the rhombohedral trilayer. 
It is possible that superconductivity in graphene bilayers appears once the pair breaking effect of spin-orbit coupling in phases with an out-of-plane spin moment is quenched by the in-plane field. This hypothesis can be tested by measurements of the in-plane and out of plane magnetic moments as a function of in-plane field.

\textit{Note added:} While finalizing this work, we became aware of a parallel theoretical analysis predicting a transition between a valley-XY quarter metal and a valley-Ising quarter metal -- equivalent to the intervalley coherent and valley imbalanced quarter metals discussed in this work, respectively. Ref. \cite{das_quarter-metal_2023} predicts a slope of the phase transition in $B_\parallel$ in rough agreement with what we find experimentally at low in-plane fields, $\partial n^\ast / \partial B_\parallel \approx -0.5 \times 10^{11}$ cm$^{-2}$ T$^{-1}$.

\textbf{Acknowledgements.} 
The authors would like to acknowledge  discussions with L. Levitov, Z. Dong, A. Macdonald, C. Huang, M. Zaletel, S. Chatterjee, O. Vafek.  
This project was primarily funded by the Department of Energy under DE-SC0020043. 
A.F.Y. acknowledges the support of the Gordon and Betty Moore Foundation under award GBMF9471 and the Packard Foundation under award 2016-65145 for general group activities.
This research was supported in part by the National Science Foundation under Grants No. NSF PHY-1748958 and PHY-2309135.
A portion of this work was performed in the UCSB Nanofabrication Facility, an open access laboratory. 
This work made use of facilities funded by Enabling Quantum Leap: Convergent Accelerated Discovery Foundries for Quantum Materials Science, Engineering and Information (Q-AMASE-i) award number DMR-1906325 from the National Science Foundation. 
K.W. and T.T. acknowledge support from the Elemental Strategy Initiative conducted by the MEXT, Japan (Grant Number JPMXP0112101001) and JSPSKAKENHI (Grant Numbers 19H05790, 20H00354 and 21H05233). 
E.B. was supported by the European Research Council (ERC) under grant HQMAT (Grant Agreement No. 817799), the Israel-US Binational Science Foundation (BSF), and by CRC 183 of the Deutsche Forschungsgemeinschaft (project C02).  
CLT acknowledges support from the Hertz Foundation and from the National Science Foundation Graduate Research Fellowship Program under grant 1650114.

\textbf{Author Contributions.} 
A.F.Y. conceived of and directed the project.
H.Z. and T.X. designed and fabricated the device.
C.L.T., T.A., O.S., G.B. and E.R. fabricated nanoSQUID tips for the measurements.
K.W. and T.T. grew the hBN crystals.
M.E.H., C.L.T., T.A., O.S. and E.R. developed the nanoSQUID microscope.
T.A., O.S., C.L.T. performed nanoSQUID measurements and T.A., O.S., H.Z., H.M.Y., C.L.P., and L.H. performed capacitance measurements.
J.X., Y.V., T.H., and E. B. contributed to the theoretical interpretation and performed the numerical simulations.
T.A., O.S., E.B., and A.F.Y. wrote the paper with inputs from all other authors.

\section{Materials and Methods}

\begin{figure}[ht]
    \centering
    \includegraphics{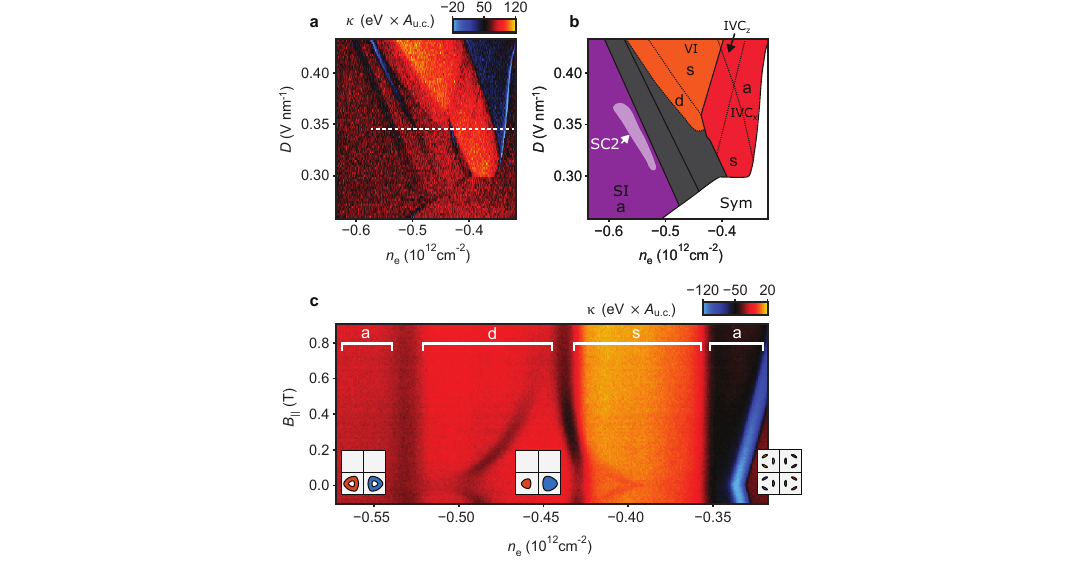}
    \caption{\textbf{Electronic Phases in the vicinity of SC2}
     (\textbf{A}) Inverse compressibilty as a function of $B_\perp$ and $n_e$ in the vicinity of the quarter metal at $B_\perp$ = 0.
    (\textbf{B}) Schematic phase diagram indicating the electronic phases of \textbf{A} as a function of $n_e$. Solid lines between different colored areas represent first order phase transitions in $\kappa$ between electronic phases: Sym, VI, IVC and SI. Dashed lines indicate Lifshitz transitions between disjointed (d), simple (s) and annular (a) Fermi surfaces as marked by lower case labels. There is a superconducting pocket (SC2) at the boundary between the IVC and SI phases as marked.\cite{zhou_superconductivity_2021}
    (\textbf{C}) Inverse compressibility, $\kappa$, as a function of carrier density $n_e$ and versus $B_{\parallel}$ along the dashed line in \textbf{A}.
    \figurelabel{fig:ext\_SC2}
    }
    \label{fig:ext_SC2}
\end{figure}

\begin{figure}[ht]
    \centering
    \includegraphics{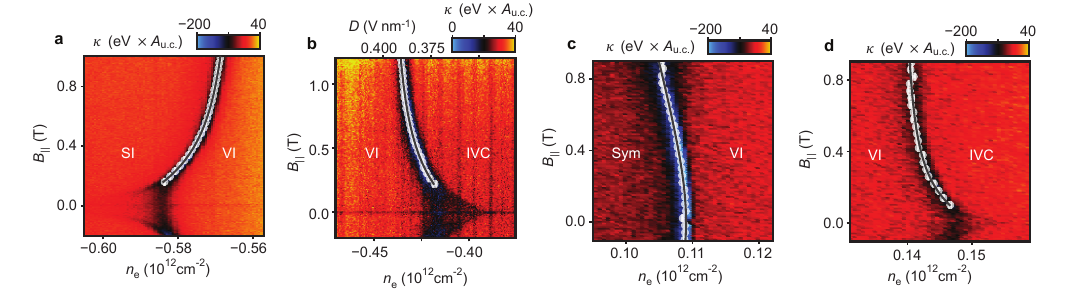}
    \caption{\textbf{Estimation of $\lambda$ from different transitions.}
     (\textbf{A}) SI to VI transition at hole doping, at $D = 0.46\,\text{V}\,\text{nm}^{-1}$.  White points represent extracted transition location; the dark line indicates the fit described in the Methods. 
     (\textbf{B}) fit for the VI to IVC transition at hole doping,
     (\textbf{C}) fit for the Sym to VI transition at electron doping, for $D = -0.23\,\text{V}\,\text{nm}^{-1}$, and
     (\textbf{D}) VI to IVC transition for electron doping for $D = -0.23\,\text{V}\,\text{nm}^{-1}$.
     The extracted $(\lambda_{min},\lambda_{fit},\lambda_{max})$, in $\mu eV$, for the transitions are A, $(25,46,60)$; B, $(39,56,98)$; C, $(54,65,174)$; D, $(29,32,52)$. 
     \figurelabel{fig:ext\_data\_electron}
    }
    \label{fig:ext_data_electron}
\end{figure}

\paragraph{\textbf{Compressibility and electronic transport measurements:}}\label{sec:capactiance}
Ref.~\cite{zhou_half-_2021} and ref.~\cite{zhou_superconductivity_2021} studied the same device, and details of the sample fabrication, electronic transport, and compressibility measurements are provided in those references. In contrast to those references, however, data in Figs. \ref{fig:1}B,E, \ref{fig:3}A,E,F,G, \ref{fig:4}A,C,D, \ref{fig:ext_data_electron}, \ref{fig:ext_SC2} were taken in an improved compressibility set-up that introduced a decoupling capacitor between the transistor amplifier and the sample as well as a second stage cryogenic amplifier located at the 4K stage.  This setup resulted in lower electronic temperature and provided improved signal to noise, allowing for smaller amplitude modulation and consequently better resolution of fine features (see Ref. \cite{holleis_ising_2023} for details).    

\paragraph{\textbf{nanoSQUID on tip measurements:}}\label{sec:nSOT}
The nanoSQUID on tip (nSOT) microscopy was performed with an indium SQUID fabricated at the tip of a pulled quartz pipette\cite{anahory_squid--tip_2020}.  The field period of 71 mT corresponds to an effective diameter of 193 nm. 
The nSOT signal was acquired using a series SQUID array amplifier (SSAA) in feedback mode, where the feedback voltage $V_\text{SSAA}$ is proportional to the current through the nSOT\cite{anahory_squid--tip_2020}. 
The SQUID was mechanically controlled with a piezoelectrically pumped quartz tuning fork in a phase locked loop and positioned over the sample by piezoelectric nanopositioners. Unless otherwise noted the magnetic data was taken in constant height mode 150 nm above the surface of the sample, corresponding to a height $h \approx 300$~nm above the ABC graphene layer. 

\begin{figure*}[ht]
    \centering
    \includegraphics[width = 130mm]{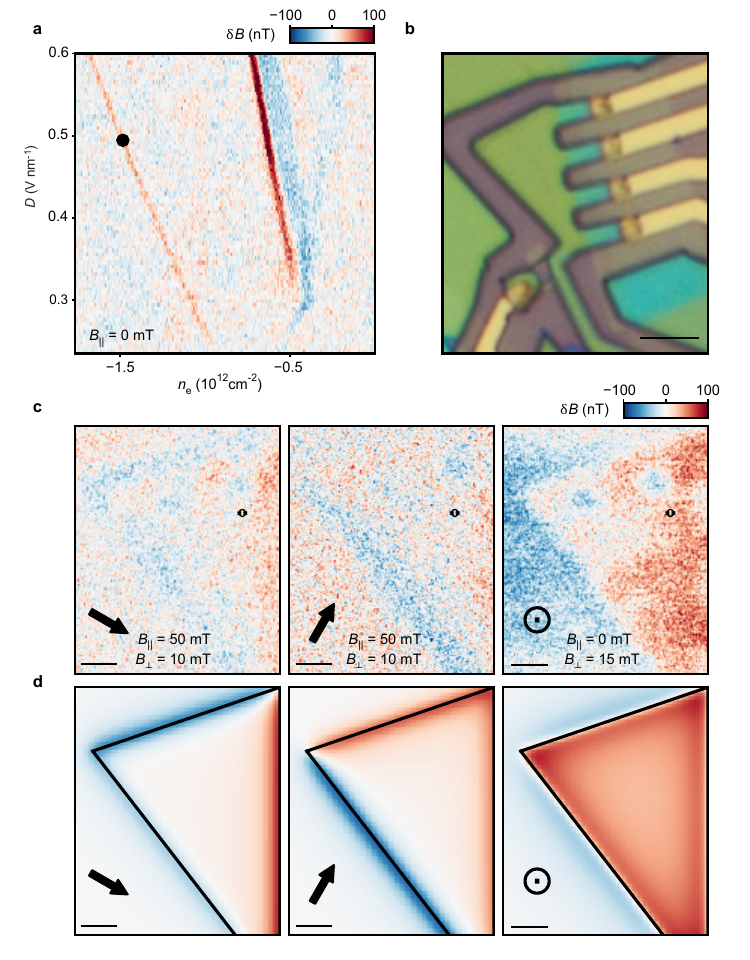}
    \caption{\textbf{Half metal ferromagnetism.}
    (\textbf{A}) The nSOT phase diagram of the hole side versus density and displacement field. Most nSOT data in this work focuses on magnetism in the vicinity of the VI phase, in this figure we focus on magnetism near the SI transition, marked by black circle. 
    (\textbf{B}) Optical image of device. Scale bar is 3~$\mu$m.
    (\textbf{C}) Spatial images of $\delta B$ for different directions of in-plane and out-of-plane field with $n_{e}$ and D set on the half metal transition, marked with a back circle in A. Marked location indicates where the the data in Fig.~\ref{fig:1}D was taken. Scale bars are 1~$\mu$m.
    (\textbf{D}) Simulated magnetism from a sheet of dipoles in the shape of the sample with magnetic moment oriented in the direction of the fields in D and the signal normalized to correspond to the total density $n_{e}$ with each carrier having $1 \mu_{B}$ of magnetic moment. The qualitative and approximate quantitative agreement of the data in D and the simple dipole model in E indicates that magnetism arises due to the spins of the carriers aligning with the external field. Scale bars are 1~$\mu$m.
    \figurelabel{fig:ext\_data\_half\_metal}
    }
    \label{fig:ext_data_half_metal}
\end{figure*}

At 300mK, the nSOT had good sensitivity up to $H_{c,\perp} \geq 500$~mT. During operation the magnetic field and nSOT voltage bias are held constant at a particular working point determined by the applied bias and magnetic field, with the bias selected to maximize the sensitivity at a given field. For the measurements shown here, the sensitivity varied between 25-35 nT/$\sqrt{\mathrm{Hz}}$.  A small out-of-plane field is typically necessary to get reasonable sensitivity:  for Fig.~\ref{fig:1}C and \ref{fig:ext_data_half_metal}A, $(B_\perp,B_\parallel)=(15 \mathrm{mT}, 0)$, and for Fig.~\ref{fig:1}D,E $(B_\perp,B_\parallel)=(19 \mathrm{mT}, 500 \mathrm{mT})$.

As described in the main text, magnetometry measurements are performed by applying a finite frequency (typically 0.5-4 kHz, chosen to minimize noise) excitation to the bottom gate (with peak-peak amplitude $15$~mV for Fig.~\ref{fig:1}B and $35$~mV otherwise)  and measuring the SQUID response at the same frequency. Images thus show contrast at the onset (and disappearance) of magnetic phases in the $n_{e}$-$D$ plane. 
Fig.~\ref{fig:1}C shows a real-space image of the device acquired at  $n_{e} = -0.67 \times 10^{12}$~cm$^{-1}$  and $D=0.55$~V/nm. 

\paragraph{\textbf{Determination of $\lambda$}:}

In the main text Fig.~\ref{fig:3}A and Extended Data Fig.~\ref{fig:ext_data_electron} we fit two different functional forms for the curvature of spin-orbit affected transitions as a function of $B_\parallel$. Three free parameters were included in the fit, labeled below as $n_\text{offset}$, $\alpha$, and $\lambda$. 
\begin{widetext}
\begin{equation}
n_e^\ast(B_\parallel, n_\text{offset}, \alpha, \lambda)=\left\{\begin{array}{ll}
   n_\text{offset} + \alpha \left ( B_\parallel+\lambda/g\mu_B-\sqrt{B_\parallel^2+(\lambda/g\mu_B)^2} \right)   & \text{Fig.~\ref{fig:3}A, Fig.~\ref{fig:ext_data_electron}A,B,D} \\
   n_\text{offset} + \alpha \left ( \lambda/g\mu_B-\sqrt{B_\parallel^2+(\lambda/g\mu_B)^2} \right)   & \text{Fig.~\ref{fig:ext_data_electron}C}
\end{array}\right.
\end{equation}
\end{widetext}

Here, $n_\text{offset}$ represents a net density offset to account for the finite density location of transitions, $\lambda$ is the SO coupling strength, and $\alpha = \pm\frac{|m_{spin}|}{\Delta\mu}$.  
A complete derivation of the functional forms shown here can be found in the supplementary information. 
The location of the transition  was determined by fitting compressibility data to a Gaussian peak.
In Fig.~\ref{fig:ext_data_electron}F, we give a complete list of extracted $\lambda$ values for all four studied transitions. Near zero field our data is affected by the presence of additional phases, particularly the IVC$_\mathrm{z}$, which prevents effective determination of $n_e^\ast$. At high $B_\parallel$, meanwhile, the transitions saturate to a fixed $n^*_e$, and fits become insensitive to $\lambda$.  The dominant source of error thus arises from the choice of range for the fit. 
To capture this uncertainty, we fit both the whole range as well as all sub-ranges between the  high- and low $B_\parallel$ limits that span at least fifteen times the single-point error in $n^*_e$ as determined from the standard deviation of the Gaussian peak fit. 
In Extended Data Fig. ~\ref{fig:ext_data_electron}, we report both the best fit to the whole curve as well as the  maximum and minimum fits from the sub-ranges, $\lambda_\text{max}$ and $\lambda_\text{min}$. The best fit for the four transitions give values between 30-65 $\mu$eV.  

\paragraph{\textbf{Determination of $\Delta m$:}}\label{sec:m_determination}

Knowing $\Delta \mu$ at a phase boundary allows, in principle, for precise determination of $\Delta m$. To measure $\Delta \mu$, we integrate our thermodynamic compressibility $\kappa=d\mu/dn_e$ over the density range of the negative $\kappa$ peak corresponding to a phase transition. In practice,  integrating our experimental data is challenging due to the weakness of the first order transitions and differences between the compressibility of the two adjoining phases. We thus expect an overall systematic error; for example, systematic underestimates of $\Delta \mu$ may be responsible for the smaller than expected values of $\Delta m_\parallel$ reported in Fig. \ref{fig:3}B.  However, we do not expect this type of systematic error to be magnetic field dependent, allowing comparison of the field dependence of $\Delta \mu$ and $\partial n^\ast/\partial B$.  This comparison is crucial to determining whether the curvature observed in a phase transition arises from a change in $\Delta m$--curvature may also be generated by a field-dependent $\Delta \mu$. 

In the data shown in Fig.~\ref{fig:3}B, we integrate compressibility data to produce the plotted $\Delta \mu$.  This data shows no systematic trend above our noise floor, so we assume a constant $\Delta \mu=$ over the depicted range, corresponding to the average measured value.  
 $\partial n^\ast / \partial B$ is determined by numerically differentiating the $n_e^*$ values extracted from peak fitting as described in the previous section. 
 $\Delta m_\parallel$ is then calculated by multiplying this number by the average $\Delta \mu$. 

\bibliographystyle{apsrev4-2}
\bibliography{references}

\begin{thebibliography}{59}%
\makeatletter
\providecommand \@ifxundefined [1]{%
 \@ifx{#1\undefined}
}%
\providecommand \@ifnum [1]{%
 \ifnum #1\expandafter \@firstoftwo
 \else \expandafter \@secondoftwo
 \fi
}%
\providecommand \@ifx [1]{%
 \ifx #1\expandafter \@firstoftwo
 \else \expandafter \@secondoftwo
 \fi
}%
\providecommand \natexlab [1]{#1}%
\providecommand \enquote  [1]{``#1''}%
\providecommand \bibnamefont  [1]{#1}%
\providecommand \bibfnamefont [1]{#1}%
\providecommand \citenamefont [1]{#1}%
\providecommand \href@noop [0]{\@secondoftwo}%
\providecommand \href [0]{\begingroup \@sanitize@url \@href}%
\providecommand \@href[1]{\@@startlink{#1}\@@href}%
\providecommand \@@href[1]{\endgroup#1\@@endlink}%
\providecommand \@sanitize@url [0]{\catcode `\\12\catcode `\$12\catcode
  `\&12\catcode `\#12\catcode `\^12\catcode `\_12\catcode `\%12\relax}%
\providecommand \@@startlink[1]{}%
\providecommand \@@endlink[0]{}%
\providecommand \url  [0]{\begingroup\@sanitize@url \@url }%
\providecommand \@url [1]{\endgroup\@href {#1}{\urlprefix }}%
\providecommand \urlprefix  [0]{URL }%
\providecommand \Eprint [0]{\href }%
\providecommand \doibase [0]{https://doi.org/}%
\providecommand \selectlanguage [0]{\@gobble}%
\providecommand \bibinfo  [0]{\@secondoftwo}%
\providecommand \bibfield  [0]{\@secondoftwo}%
\providecommand \translation [1]{[#1]}%
\providecommand \BibitemOpen [0]{}%
\providecommand \bibitemStop [0]{}%
\providecommand \bibitemNoStop [0]{.\EOS\space}%
\providecommand \EOS [0]{\spacefactor3000\relax}%
\providecommand \BibitemShut  [1]{\csname bibitem#1\endcsname}%
\let\auto@bib@innerbib\@empty
\bibitem [{\citenamefont {Zhou}\ \emph
  {et~al.}(2021{\natexlab{a}})\citenamefont {Zhou}, \citenamefont {Xie},
  \citenamefont {Taniguchi}, \citenamefont {Watanabe},\ and\ \citenamefont
  {Young}}]{zhou_superconductivity_2021}%
  \BibitemOpen
  \bibfield  {author} {\bibinfo {author} {\bibfnamefont {H.}~\bibnamefont
  {Zhou}}, \bibinfo {author} {\bibfnamefont {T.}~\bibnamefont {Xie}}, \bibinfo
  {author} {\bibfnamefont {T.}~\bibnamefont {Taniguchi}}, \bibinfo {author}
  {\bibfnamefont {K.}~\bibnamefont {Watanabe}},\ and\ \bibinfo {author}
  {\bibfnamefont {A.~F.}\ \bibnamefont {Young}},\ }\href
  {https://doi.org/10.1038/s41586-021-03926-0} {\bibfield  {journal} {\bibinfo
  {journal} {Nature}\ }\textbf {\bibinfo {volume} {598}},\ \bibinfo {pages}
  {434} (\bibinfo {year} {2021}{\natexlab{a}})}\BibitemShut {NoStop}%
\bibitem [{\citenamefont {Zhou}\ \emph {et~al.}(2022)\citenamefont {Zhou},
  \citenamefont {Holleis}, \citenamefont {Saito}, \citenamefont {Cohen},
  \citenamefont {Huynh}, \citenamefont {Patterson}, \citenamefont {Yang},
  \citenamefont {Taniguchi}, \citenamefont {Watanabe},\ and\ \citenamefont
  {Young}}]{zhou_isospin_2022}%
  \BibitemOpen
  \bibfield  {author} {\bibinfo {author} {\bibfnamefont {H.}~\bibnamefont
  {Zhou}}, \bibinfo {author} {\bibfnamefont {L.}~\bibnamefont {Holleis}},
  \bibinfo {author} {\bibfnamefont {Y.}~\bibnamefont {Saito}}, \bibinfo
  {author} {\bibfnamefont {L.}~\bibnamefont {Cohen}}, \bibinfo {author}
  {\bibfnamefont {W.}~\bibnamefont {Huynh}}, \bibinfo {author} {\bibfnamefont
  {C.~L.}\ \bibnamefont {Patterson}}, \bibinfo {author} {\bibfnamefont
  {F.}~\bibnamefont {Yang}}, \bibinfo {author} {\bibfnamefont {T.}~\bibnamefont
  {Taniguchi}}, \bibinfo {author} {\bibfnamefont {K.}~\bibnamefont
  {Watanabe}},\ and\ \bibinfo {author} {\bibfnamefont {A.~F.}\ \bibnamefont
  {Young}},\ }\href {https://doi.org/10.1126/science.abm8386} {\bibfield
  {journal} {\bibinfo  {journal} {Science}\ }\textbf {\bibinfo {volume}
  {375}},\ \bibinfo {pages} {774} (\bibinfo {year} {2022})}\BibitemShut
  {NoStop}%
\bibitem [{\citenamefont {Zhang}\ \emph {et~al.}(2023)\citenamefont {Zhang},
  \citenamefont {Polski}, \citenamefont {Thomson}, \citenamefont
  {Lantagne-Hurtubise}, \citenamefont {Lewandowski}, \citenamefont {Zhou},
  \citenamefont {Watanabe}, \citenamefont {Taniguchi}, \citenamefont {Alicea},\
  and\ \citenamefont {Nadj-Perge}}]{zhang_enhanced_2023}%
  \BibitemOpen
  \bibfield  {author} {\bibinfo {author} {\bibfnamefont {Y.}~\bibnamefont
  {Zhang}}, \bibinfo {author} {\bibfnamefont {R.}~\bibnamefont {Polski}},
  \bibinfo {author} {\bibfnamefont {A.}~\bibnamefont {Thomson}}, \bibinfo
  {author} {\bibfnamefont {E.}~\bibnamefont {Lantagne-Hurtubise}}, \bibinfo
  {author} {\bibfnamefont {C.}~\bibnamefont {Lewandowski}}, \bibinfo {author}
  {\bibfnamefont {H.}~\bibnamefont {Zhou}}, \bibinfo {author} {\bibfnamefont
  {K.}~\bibnamefont {Watanabe}}, \bibinfo {author} {\bibfnamefont
  {T.}~\bibnamefont {Taniguchi}}, \bibinfo {author} {\bibfnamefont
  {J.}~\bibnamefont {Alicea}},\ and\ \bibinfo {author} {\bibfnamefont
  {S.}~\bibnamefont {Nadj-Perge}},\ }\href
  {https://doi.org/10.1038/s41586-022-05446-x} {\bibfield  {journal} {\bibinfo
  {journal} {Nature}\ }\textbf {\bibinfo {volume} {613}},\ \bibinfo {pages}
  {268} (\bibinfo {year} {2023})}\BibitemShut {NoStop}%
\bibitem [{\citenamefont {Holleis}\ \emph {et~al.}(2023)\citenamefont
  {Holleis}, \citenamefont {Patterson}, \citenamefont {Zhang}, \citenamefont
  {Yoo}, \citenamefont {Zhou}, \citenamefont {Taniguchi}, \citenamefont
  {Watanabe}, \citenamefont {Nadj-Perge},\ and\ \citenamefont
  {Young}}]{holleis_ising_2023}%
  \BibitemOpen
  \bibfield  {author} {\bibinfo {author} {\bibfnamefont {L.}~\bibnamefont
  {Holleis}}, \bibinfo {author} {\bibfnamefont {C.~L.}\ \bibnamefont
  {Patterson}}, \bibinfo {author} {\bibfnamefont {Y.}~\bibnamefont {Zhang}},
  \bibinfo {author} {\bibfnamefont {H.~M.}\ \bibnamefont {Yoo}}, \bibinfo
  {author} {\bibfnamefont {H.}~\bibnamefont {Zhou}}, \bibinfo {author}
  {\bibfnamefont {T.}~\bibnamefont {Taniguchi}}, \bibinfo {author}
  {\bibfnamefont {K.}~\bibnamefont {Watanabe}}, \bibinfo {author}
  {\bibfnamefont {S.}~\bibnamefont {Nadj-Perge}},\ and\ \bibinfo {author}
  {\bibfnamefont {A.~F.}\ \bibnamefont {Young}},\ }\href
  {https://doi.org/10.48550/arXiv.2303.00742} {\bibinfo {title} {Ising
  {Superconductivity} and {Nematicity} in {Bernal} {Bilayer} {Graphene} with
  {Strong} {Spin} {Orbit} {Coupling}}} (\bibinfo {year} {2023}),\ \bibinfo
  {note} {arXiv:2303.00742 [cond-mat]}\BibitemShut {NoStop}%
\bibitem [{\citenamefont {Shi}\ \emph {et~al.}(2020)\citenamefont {Shi},
  \citenamefont {Xu}, \citenamefont {Yang}, \citenamefont {Slizovskiy},
  \citenamefont {Morozov}, \citenamefont {Son}, \citenamefont {Ozdemir},
  \citenamefont {Mullan}, \citenamefont {Barrier}, \citenamefont {Yin},
  \citenamefont {Berdyugin}, \citenamefont {Piot}, \citenamefont {Taniguchi},
  \citenamefont {Watanabe}, \citenamefont {Fal’ko}, \citenamefont
  {Novoselov}, \citenamefont {Geim},\ and\ \citenamefont
  {Mishchenko}}]{shi_electronic_2020}%
  \BibitemOpen
  \bibfield  {author} {\bibinfo {author} {\bibfnamefont {Y.}~\bibnamefont
  {Shi}}, \bibinfo {author} {\bibfnamefont {S.}~\bibnamefont {Xu}}, \bibinfo
  {author} {\bibfnamefont {Y.}~\bibnamefont {Yang}}, \bibinfo {author}
  {\bibfnamefont {S.}~\bibnamefont {Slizovskiy}}, \bibinfo {author}
  {\bibfnamefont {S.~V.}\ \bibnamefont {Morozov}}, \bibinfo {author}
  {\bibfnamefont {S.-K.}\ \bibnamefont {Son}}, \bibinfo {author} {\bibfnamefont
  {S.}~\bibnamefont {Ozdemir}}, \bibinfo {author} {\bibfnamefont
  {C.}~\bibnamefont {Mullan}}, \bibinfo {author} {\bibfnamefont
  {J.}~\bibnamefont {Barrier}}, \bibinfo {author} {\bibfnamefont
  {J.}~\bibnamefont {Yin}}, \bibinfo {author} {\bibfnamefont {A.~I.}\
  \bibnamefont {Berdyugin}}, \bibinfo {author} {\bibfnamefont {B.~A.}\
  \bibnamefont {Piot}}, \bibinfo {author} {\bibfnamefont {T.}~\bibnamefont
  {Taniguchi}}, \bibinfo {author} {\bibfnamefont {K.}~\bibnamefont {Watanabe}},
  \bibinfo {author} {\bibfnamefont {V.~I.}\ \bibnamefont {Fal’ko}}, \bibinfo
  {author} {\bibfnamefont {K.~S.}\ \bibnamefont {Novoselov}}, \bibinfo {author}
  {\bibfnamefont {A.~K.}\ \bibnamefont {Geim}},\ and\ \bibinfo {author}
  {\bibfnamefont {A.}~\bibnamefont {Mishchenko}},\ }\href
  {https://doi.org/10.1038/s41586-020-2568-2} {\bibfield  {journal} {\bibinfo
  {journal} {Nature}\ }\textbf {\bibinfo {volume} {584}},\ \bibinfo {pages}
  {210} (\bibinfo {year} {2020})}\BibitemShut {NoStop}%
\bibitem [{\citenamefont {Zhou}\ \emph
  {et~al.}(2021{\natexlab{b}})\citenamefont {Zhou}, \citenamefont {Xie},
  \citenamefont {Ghazaryan}, \citenamefont {Holder}, \citenamefont {Ehrets},
  \citenamefont {Spanton}, \citenamefont {Taniguchi}, \citenamefont {Watanabe},
  \citenamefont {Berg}, \citenamefont {Serbyn},\ and\ \citenamefont
  {Young}}]{zhou_half-_2021}%
  \BibitemOpen
  \bibfield  {author} {\bibinfo {author} {\bibfnamefont {H.}~\bibnamefont
  {Zhou}}, \bibinfo {author} {\bibfnamefont {T.}~\bibnamefont {Xie}}, \bibinfo
  {author} {\bibfnamefont {A.}~\bibnamefont {Ghazaryan}}, \bibinfo {author}
  {\bibfnamefont {T.}~\bibnamefont {Holder}}, \bibinfo {author} {\bibfnamefont
  {J.~R.}\ \bibnamefont {Ehrets}}, \bibinfo {author} {\bibfnamefont {E.~M.}\
  \bibnamefont {Spanton}}, \bibinfo {author} {\bibfnamefont {T.}~\bibnamefont
  {Taniguchi}}, \bibinfo {author} {\bibfnamefont {K.}~\bibnamefont {Watanabe}},
  \bibinfo {author} {\bibfnamefont {E.}~\bibnamefont {Berg}}, \bibinfo {author}
  {\bibfnamefont {M.}~\bibnamefont {Serbyn}},\ and\ \bibinfo {author}
  {\bibfnamefont {A.~F.}\ \bibnamefont {Young}},\ }\href
  {https://doi.org/10.1038/s41586-021-03938-w} {\bibfield  {journal} {\bibinfo
  {journal} {Nature}\ }\textbf {\bibinfo {volume} {598}},\ \bibinfo {pages}
  {429} (\bibinfo {year} {2021}{\natexlab{b}})}\BibitemShut {NoStop}%
\bibitem [{\citenamefont {de~la Barrera}\ \emph {et~al.}(2022)\citenamefont
  {de~la Barrera}, \citenamefont {Aronson}, \citenamefont {Zheng},
  \citenamefont {Watanabe}, \citenamefont {Taniguchi}, \citenamefont {Ma},
  \citenamefont {Jarillo-Herrero},\ and\ \citenamefont
  {Ashoori}}]{de_la_barrera_cascade_2022}%
  \BibitemOpen
  \bibfield  {author} {\bibinfo {author} {\bibfnamefont {S.~C.}\ \bibnamefont
  {de~la Barrera}}, \bibinfo {author} {\bibfnamefont {S.}~\bibnamefont
  {Aronson}}, \bibinfo {author} {\bibfnamefont {Z.}~\bibnamefont {Zheng}},
  \bibinfo {author} {\bibfnamefont {K.}~\bibnamefont {Watanabe}}, \bibinfo
  {author} {\bibfnamefont {T.}~\bibnamefont {Taniguchi}}, \bibinfo {author}
  {\bibfnamefont {Q.}~\bibnamefont {Ma}}, \bibinfo {author} {\bibfnamefont
  {P.}~\bibnamefont {Jarillo-Herrero}},\ and\ \bibinfo {author} {\bibfnamefont
  {R.}~\bibnamefont {Ashoori}},\ }\href
  {https://doi.org/10.1038/s41567-022-01616-w} {\bibfield  {journal} {\bibinfo
  {journal} {Nature Physics}\ }\textbf {\bibinfo {volume} {18}},\ \bibinfo
  {pages} {771} (\bibinfo {year} {2022})}\BibitemShut {NoStop}%
\bibitem [{\citenamefont {Seiler}\ \emph {et~al.}(2022)\citenamefont {Seiler},
  \citenamefont {Geisenhof}, \citenamefont {Winterer}, \citenamefont
  {Watanabe}, \citenamefont {Taniguchi}, \citenamefont {Xu}, \citenamefont
  {Zhang},\ and\ \citenamefont {Weitz}}]{seiler_quantum_2022}%
  \BibitemOpen
  \bibfield  {author} {\bibinfo {author} {\bibfnamefont {A.~M.}\ \bibnamefont
  {Seiler}}, \bibinfo {author} {\bibfnamefont {F.~R.}\ \bibnamefont
  {Geisenhof}}, \bibinfo {author} {\bibfnamefont {F.}~\bibnamefont {Winterer}},
  \bibinfo {author} {\bibfnamefont {K.}~\bibnamefont {Watanabe}}, \bibinfo
  {author} {\bibfnamefont {T.}~\bibnamefont {Taniguchi}}, \bibinfo {author}
  {\bibfnamefont {T.}~\bibnamefont {Xu}}, \bibinfo {author} {\bibfnamefont
  {F.}~\bibnamefont {Zhang}},\ and\ \bibinfo {author} {\bibfnamefont {R.~T.}\
  \bibnamefont {Weitz}},\ }\href {https://doi.org/10.1038/s41586-022-04937-1}
  {\bibfield  {journal} {\bibinfo  {journal} {Nature}\ }\textbf {\bibinfo
  {volume} {608}},\ \bibinfo {pages} {298} (\bibinfo {year}
  {2022})}\BibitemShut {NoStop}%
\bibitem [{\citenamefont {Lin}\ \emph {et~al.}(2023)\citenamefont {Lin},
  \citenamefont {Wang}, \citenamefont {Zhang}, \citenamefont {Watanabe},
  \citenamefont {Taniguchi}, \citenamefont {Fu},\ and\ \citenamefont
  {Li}}]{lin_spontaneous_2023}%
  \BibitemOpen
  \bibfield  {author} {\bibinfo {author} {\bibfnamefont {J.-X.}\ \bibnamefont
  {Lin}}, \bibinfo {author} {\bibfnamefont {Y.}~\bibnamefont {Wang}}, \bibinfo
  {author} {\bibfnamefont {N.~J.}\ \bibnamefont {Zhang}}, \bibinfo {author}
  {\bibfnamefont {K.}~\bibnamefont {Watanabe}}, \bibinfo {author}
  {\bibfnamefont {T.}~\bibnamefont {Taniguchi}}, \bibinfo {author}
  {\bibfnamefont {L.}~\bibnamefont {Fu}},\ and\ \bibinfo {author}
  {\bibfnamefont {J.~I.~A.}\ \bibnamefont {Li}},\ }\href
  {https://doi.org/10.48550/arXiv.2302.04261} {\bibinfo {title} {Spontaneous
  momentum polarization and diodicity in {Bernal} bilayer graphene}} (\bibinfo
  {year} {2023}),\ \bibinfo {note} {arXiv:2302.04261 [cond-mat]}\BibitemShut
  {NoStop}%
\bibitem [{\citenamefont {Seiler}\ \emph {et~al.}(2023)\citenamefont {Seiler},
  \citenamefont {Statz}, \citenamefont {Weimer}, \citenamefont {Jacobsen},
  \citenamefont {Watanabe}, \citenamefont {Taniguchi}, \citenamefont {Dong},
  \citenamefont {Levitov},\ and\ \citenamefont
  {Weitz}}]{seiler_interaction-driven_2023}%
  \BibitemOpen
  \bibfield  {author} {\bibinfo {author} {\bibfnamefont {A.~M.}\ \bibnamefont
  {Seiler}}, \bibinfo {author} {\bibfnamefont {M.}~\bibnamefont {Statz}},
  \bibinfo {author} {\bibfnamefont {I.}~\bibnamefont {Weimer}}, \bibinfo
  {author} {\bibfnamefont {N.}~\bibnamefont {Jacobsen}}, \bibinfo {author}
  {\bibfnamefont {K.}~\bibnamefont {Watanabe}}, \bibinfo {author}
  {\bibfnamefont {T.}~\bibnamefont {Taniguchi}}, \bibinfo {author}
  {\bibfnamefont {Z.}~\bibnamefont {Dong}}, \bibinfo {author} {\bibfnamefont
  {L.~S.}\ \bibnamefont {Levitov}},\ and\ \bibinfo {author} {\bibfnamefont
  {R.~T.}\ \bibnamefont {Weitz}},\ }\href
  {https://doi.org/10.48550/arXiv.2308.00827} {\bibinfo {title}
  {Interaction-driven (quasi-) insulating ground states of gapped
  electron-doped bilayer graphene}} (\bibinfo {year} {2023}),\ \bibinfo {note}
  {arXiv:2308.00827 [cond-mat]}\BibitemShut {NoStop}%
\bibitem [{\citenamefont {Han}\ \emph {et~al.}(2023{\natexlab{a}})\citenamefont
  {Han}, \citenamefont {Lu}, \citenamefont {Scuri}, \citenamefont {Sung},
  \citenamefont {Wang}, \citenamefont {Han}, \citenamefont {Watanabe},
  \citenamefont {Taniguchi}, \citenamefont {Fu}, \citenamefont {Park},\ and\
  \citenamefont {Ju}}]{han_orbital_2023}%
  \BibitemOpen
  \bibfield  {author} {\bibinfo {author} {\bibfnamefont {T.}~\bibnamefont
  {Han}}, \bibinfo {author} {\bibfnamefont {Z.}~\bibnamefont {Lu}}, \bibinfo
  {author} {\bibfnamefont {G.}~\bibnamefont {Scuri}}, \bibinfo {author}
  {\bibfnamefont {J.}~\bibnamefont {Sung}}, \bibinfo {author} {\bibfnamefont
  {J.}~\bibnamefont {Wang}}, \bibinfo {author} {\bibfnamefont {T.}~\bibnamefont
  {Han}}, \bibinfo {author} {\bibfnamefont {K.}~\bibnamefont {Watanabe}},
  \bibinfo {author} {\bibfnamefont {T.}~\bibnamefont {Taniguchi}}, \bibinfo
  {author} {\bibfnamefont {L.}~\bibnamefont {Fu}}, \bibinfo {author}
  {\bibfnamefont {H.}~\bibnamefont {Park}},\ and\ \bibinfo {author}
  {\bibfnamefont {L.}~\bibnamefont {Ju}},\ }\href
  {https://doi.org/10.48550/arXiv.2308.08837} {\bibinfo {title} {Orbital
  {Multiferroicity} in {Pentalayer} {Rhombohedral} {Graphene}}} (\bibinfo
  {year} {2023}{\natexlab{a}}),\ \bibinfo {note} {arXiv:2308.08837
  [cond-mat]}\BibitemShut {NoStop}%
\bibitem [{\citenamefont {Kane}\ and\ \citenamefont
  {Mele}(2005)}]{kane_quantum_2005}%
  \BibitemOpen
  \bibfield  {author} {\bibinfo {author} {\bibfnamefont {C.~L.}\ \bibnamefont
  {Kane}}\ and\ \bibinfo {author} {\bibfnamefont {E.~J.}\ \bibnamefont
  {Mele}},\ }\href {https://doi.org/10.1103/PhysRevLett.95.226801} {\bibfield
  {journal} {\bibinfo  {journal} {Physical Review Letters}\ }\textbf {\bibinfo
  {volume} {95}},\ \bibinfo {pages} {226801} (\bibinfo {year}
  {2005})}\BibitemShut {NoStop}%
\bibitem [{\citenamefont {Min}\ \emph {et~al.}(2006)\citenamefont {Min},
  \citenamefont {Hill}, \citenamefont {Sinitsyn}, \citenamefont {Sahu},
  \citenamefont {Kleinman},\ and\ \citenamefont
  {MacDonald}}]{min_intrinsic_2006}%
  \BibitemOpen
  \bibfield  {author} {\bibinfo {author} {\bibfnamefont {H.}~\bibnamefont
  {Min}}, \bibinfo {author} {\bibfnamefont {J.~E.}\ \bibnamefont {Hill}},
  \bibinfo {author} {\bibfnamefont {N.~A.}\ \bibnamefont {Sinitsyn}}, \bibinfo
  {author} {\bibfnamefont {B.~R.}\ \bibnamefont {Sahu}}, \bibinfo {author}
  {\bibfnamefont {L.}~\bibnamefont {Kleinman}},\ and\ \bibinfo {author}
  {\bibfnamefont {A.~H.}\ \bibnamefont {MacDonald}},\ }\href
  {https://doi.org/10.1103/PhysRevB.74.165310} {\bibfield  {journal} {\bibinfo
  {journal} {Physical Review B}\ }\textbf {\bibinfo {volume} {74}},\ \bibinfo
  {pages} {165310} (\bibinfo {year} {2006})}\BibitemShut {NoStop}%
\bibitem [{\citenamefont {Yao}\ \emph {et~al.}(2007)\citenamefont {Yao},
  \citenamefont {Ye}, \citenamefont {Qi}, \citenamefont {Zhang},\ and\
  \citenamefont {Fang}}]{yao_spin-orbit_2007}%
  \BibitemOpen
  \bibfield  {author} {\bibinfo {author} {\bibfnamefont {Y.}~\bibnamefont
  {Yao}}, \bibinfo {author} {\bibfnamefont {F.}~\bibnamefont {Ye}}, \bibinfo
  {author} {\bibfnamefont {X.-L.}\ \bibnamefont {Qi}}, \bibinfo {author}
  {\bibfnamefont {S.-C.}\ \bibnamefont {Zhang}},\ and\ \bibinfo {author}
  {\bibfnamefont {Z.}~\bibnamefont {Fang}},\ }\href
  {https://doi.org/10.1103/PhysRevB.75.041401} {\bibfield  {journal} {\bibinfo
  {journal} {Physical Review B}\ }\textbf {\bibinfo {volume} {75}},\ \bibinfo
  {pages} {041401} (\bibinfo {year} {2007})}\BibitemShut {NoStop}%
\bibitem [{\citenamefont {Konschuh}\ \emph {et~al.}(2012)\citenamefont
  {Konschuh}, \citenamefont {Gmitra}, \citenamefont {Kochan},\ and\
  \citenamefont {Fabian}}]{konschuh_theory_2012}%
  \BibitemOpen
  \bibfield  {author} {\bibinfo {author} {\bibfnamefont {S.}~\bibnamefont
  {Konschuh}}, \bibinfo {author} {\bibfnamefont {M.}~\bibnamefont {Gmitra}},
  \bibinfo {author} {\bibfnamefont {D.}~\bibnamefont {Kochan}},\ and\ \bibinfo
  {author} {\bibfnamefont {J.}~\bibnamefont {Fabian}},\ }\href
  {https://doi.org/10.1103/PhysRevB.85.115423} {\bibfield  {journal} {\bibinfo
  {journal} {Physical Review B}\ }\textbf {\bibinfo {volume} {85}},\ \bibinfo
  {pages} {115423} (\bibinfo {year} {2012})}\BibitemShut {NoStop}%
\bibitem [{\citenamefont {Cvetkovic}\ \emph {et~al.}(2012)\citenamefont
  {Cvetkovic}, \citenamefont {Throckmorton},\ and\ \citenamefont
  {Vafek}}]{cvetkovic_electronic_2012}%
  \BibitemOpen
  \bibfield  {author} {\bibinfo {author} {\bibfnamefont {V.}~\bibnamefont
  {Cvetkovic}}, \bibinfo {author} {\bibfnamefont {R.~E.}\ \bibnamefont
  {Throckmorton}},\ and\ \bibinfo {author} {\bibfnamefont {O.}~\bibnamefont
  {Vafek}},\ }\href {https://doi.org/10.1103/PhysRevB.86.075467} {\bibfield
  {journal} {\bibinfo  {journal} {Physical Review B}\ }\textbf {\bibinfo
  {volume} {86}},\ \bibinfo {pages} {075467} (\bibinfo {year}
  {2012})}\BibitemShut {NoStop}%
\bibitem [{\citenamefont {Mayorov}\ \emph {et~al.}(2011)\citenamefont
  {Mayorov}, \citenamefont {Elias}, \citenamefont {Mucha-Kruczynski},
  \citenamefont {Gorbachev}, \citenamefont {Tudorovskiy}, \citenamefont
  {Zhukov}, \citenamefont {Morozov}, \citenamefont {Katsnelson}, \citenamefont
  {Geim},\ and\ \citenamefont {Novoselov}}]{mayorov_interaction-driven_2011}%
  \BibitemOpen
  \bibfield  {author} {\bibinfo {author} {\bibfnamefont {A.~S.}\ \bibnamefont
  {Mayorov}}, \bibinfo {author} {\bibfnamefont {D.~C.}\ \bibnamefont {Elias}},
  \bibinfo {author} {\bibfnamefont {M.}~\bibnamefont {Mucha-Kruczynski}},
  \bibinfo {author} {\bibfnamefont {R.~V.}\ \bibnamefont {Gorbachev}}, \bibinfo
  {author} {\bibfnamefont {T.}~\bibnamefont {Tudorovskiy}}, \bibinfo {author}
  {\bibfnamefont {A.}~\bibnamefont {Zhukov}}, \bibinfo {author} {\bibfnamefont
  {S.~V.}\ \bibnamefont {Morozov}}, \bibinfo {author} {\bibfnamefont {M.~I.}\
  \bibnamefont {Katsnelson}}, \bibinfo {author} {\bibfnamefont {A.~K.}\
  \bibnamefont {Geim}},\ and\ \bibinfo {author} {\bibfnamefont {K.~S.}\
  \bibnamefont {Novoselov}},\ }\href {https://doi.org/10.1126/science.1208683}
  {\bibfield  {journal} {\bibinfo  {journal} {Science}\ }\textbf {\bibinfo
  {volume} {333}},\ \bibinfo {pages} {860} (\bibinfo {year}
  {2011})}\BibitemShut {NoStop}%
\bibitem [{\citenamefont {Winterer}\ \emph {et~al.}(2023)\citenamefont
  {Winterer}, \citenamefont {Geisenhof}, \citenamefont {Fernandez},
  \citenamefont {Seiler}, \citenamefont {Zhang},\ and\ \citenamefont
  {Weitz}}]{winterer_ferroelectric_2023}%
  \BibitemOpen
  \bibfield  {author} {\bibinfo {author} {\bibfnamefont {F.}~\bibnamefont
  {Winterer}}, \bibinfo {author} {\bibfnamefont {F.~R.}\ \bibnamefont
  {Geisenhof}}, \bibinfo {author} {\bibfnamefont {N.}~\bibnamefont
  {Fernandez}}, \bibinfo {author} {\bibfnamefont {A.~M.}\ \bibnamefont
  {Seiler}}, \bibinfo {author} {\bibfnamefont {F.}~\bibnamefont {Zhang}},\ and\
  \bibinfo {author} {\bibfnamefont {R.~T.}\ \bibnamefont {Weitz}},\ }\href
  {https://doi.org/10.48550/arXiv.2305.04950} {\bibinfo {title} {Ferroelectric
  and anomalous quantum {Hall} states in bare rhombohedral trilayer graphene}}
  (\bibinfo {year} {2023}),\ \bibinfo {note} {arXiv:2305.04950
  [cond-mat]}\BibitemShut {NoStop}%
\bibitem [{\citenamefont {Han}\ \emph {et~al.}(2023{\natexlab{b}})\citenamefont
  {Han}, \citenamefont {Lu}, \citenamefont {Scuri}, \citenamefont {Sung},
  \citenamefont {Wang}, \citenamefont {Han}, \citenamefont {Watanabe},
  \citenamefont {Taniguchi}, \citenamefont {Park},\ and\ \citenamefont
  {Ju}}]{han_correlated_2023}%
  \BibitemOpen
  \bibfield  {author} {\bibinfo {author} {\bibfnamefont {T.}~\bibnamefont
  {Han}}, \bibinfo {author} {\bibfnamefont {Z.}~\bibnamefont {Lu}}, \bibinfo
  {author} {\bibfnamefont {G.}~\bibnamefont {Scuri}}, \bibinfo {author}
  {\bibfnamefont {J.}~\bibnamefont {Sung}}, \bibinfo {author} {\bibfnamefont
  {J.}~\bibnamefont {Wang}}, \bibinfo {author} {\bibfnamefont {T.}~\bibnamefont
  {Han}}, \bibinfo {author} {\bibfnamefont {K.}~\bibnamefont {Watanabe}},
  \bibinfo {author} {\bibfnamefont {T.}~\bibnamefont {Taniguchi}}, \bibinfo
  {author} {\bibfnamefont {H.}~\bibnamefont {Park}},\ and\ \bibinfo {author}
  {\bibfnamefont {L.}~\bibnamefont {Ju}},\ }\href
  {https://doi.org/10.48550/arXiv.2305.03151} {\bibinfo {title} {Correlated
  {Insulator} and {Chern} {Insulators} in {Pentalayer} {Rhombohedral} {Stacked}
  {Graphene}}} (\bibinfo {year} {2023}{\natexlab{b}}),\ \bibinfo {note}
  {arXiv:2305.03151 [cond-mat]}\BibitemShut {NoStop}%
\bibitem [{\citenamefont {Island}\ \emph {et~al.}(2019)\citenamefont {Island},
  \citenamefont {Cui}, \citenamefont {Lewandowski}, \citenamefont {Khoo},
  \citenamefont {Spanton}, \citenamefont {Zhou}, \citenamefont {Rhodes},
  \citenamefont {Hone}, \citenamefont {Taniguchi}, \citenamefont {Watanabe},
  \citenamefont {Levitov}, \citenamefont {Zaletel},\ and\ \citenamefont
  {Young}}]{island_spinorbit-driven_2019}%
  \BibitemOpen
  \bibfield  {author} {\bibinfo {author} {\bibfnamefont {J.~O.}\ \bibnamefont
  {Island}}, \bibinfo {author} {\bibfnamefont {X.}~\bibnamefont {Cui}},
  \bibinfo {author} {\bibfnamefont {C.}~\bibnamefont {Lewandowski}}, \bibinfo
  {author} {\bibfnamefont {J.~Y.}\ \bibnamefont {Khoo}}, \bibinfo {author}
  {\bibfnamefont {E.~M.}\ \bibnamefont {Spanton}}, \bibinfo {author}
  {\bibfnamefont {H.}~\bibnamefont {Zhou}}, \bibinfo {author} {\bibfnamefont
  {D.}~\bibnamefont {Rhodes}}, \bibinfo {author} {\bibfnamefont {J.~C.}\
  \bibnamefont {Hone}}, \bibinfo {author} {\bibfnamefont {T.}~\bibnamefont
  {Taniguchi}}, \bibinfo {author} {\bibfnamefont {K.}~\bibnamefont {Watanabe}},
  \bibinfo {author} {\bibfnamefont {L.~S.}\ \bibnamefont {Levitov}}, \bibinfo
  {author} {\bibfnamefont {M.~P.}\ \bibnamefont {Zaletel}},\ and\ \bibinfo
  {author} {\bibfnamefont {A.~F.}\ \bibnamefont {Young}},\ }\href
  {https://doi.org/10.1038/s41586-019-1304-2} {\bibfield  {journal} {\bibinfo
  {journal} {Nature}\ }\textbf {\bibinfo {volume} {571}},\ \bibinfo {pages}
  {85} (\bibinfo {year} {2019})}\BibitemShut {NoStop}%
\bibitem [{\citenamefont {You}\ and\ \citenamefont
  {Vishwanath}(2022)}]{you_kohn-luttinger_2022}%
  \BibitemOpen
  \bibfield  {author} {\bibinfo {author} {\bibfnamefont {Y.-Z.}\ \bibnamefont
  {You}}\ and\ \bibinfo {author} {\bibfnamefont {A.}~\bibnamefont
  {Vishwanath}},\ }\href {https://doi.org/10.1103/PhysRevB.105.134524}
  {\bibfield  {journal} {\bibinfo  {journal} {Physical Review B}\ }\textbf
  {\bibinfo {volume} {105}},\ \bibinfo {pages} {134524} (\bibinfo {year}
  {2022})}\BibitemShut {NoStop}%
\bibitem [{\citenamefont {Ghazaryan}\ \emph {et~al.}(2021)\citenamefont
  {Ghazaryan}, \citenamefont {Holder}, \citenamefont {Serbyn},\ and\
  \citenamefont {Berg}}]{ghazaryan_unconventional_2021}%
  \BibitemOpen
  \bibfield  {author} {\bibinfo {author} {\bibfnamefont {A.}~\bibnamefont
  {Ghazaryan}}, \bibinfo {author} {\bibfnamefont {T.}~\bibnamefont {Holder}},
  \bibinfo {author} {\bibfnamefont {M.}~\bibnamefont {Serbyn}},\ and\ \bibinfo
  {author} {\bibfnamefont {E.}~\bibnamefont {Berg}},\ }\href
  {https://doi.org/10.1103/PhysRevLett.127.247001} {\bibfield  {journal}
  {\bibinfo  {journal} {Physical Review Letters}\ }\textbf {\bibinfo {volume}
  {127}},\ \bibinfo {pages} {247001} (\bibinfo {year} {2021})}\BibitemShut
  {NoStop}%
\bibitem [{\citenamefont {Ghazaryan}\ \emph {et~al.}(2023)\citenamefont
  {Ghazaryan}, \citenamefont {Holder}, \citenamefont {Berg},\ and\
  \citenamefont {Serbyn}}]{ghazaryan_multilayer_2023}%
  \BibitemOpen
  \bibfield  {author} {\bibinfo {author} {\bibfnamefont {A.}~\bibnamefont
  {Ghazaryan}}, \bibinfo {author} {\bibfnamefont {T.}~\bibnamefont {Holder}},
  \bibinfo {author} {\bibfnamefont {E.}~\bibnamefont {Berg}},\ and\ \bibinfo
  {author} {\bibfnamefont {M.}~\bibnamefont {Serbyn}},\ }\href
  {https://doi.org/10.1103/PhysRevB.107.104502} {\bibfield  {journal} {\bibinfo
   {journal} {Physical Review B}\ }\textbf {\bibinfo {volume} {107}},\ \bibinfo
  {pages} {104502} (\bibinfo {year} {2023})}\BibitemShut {NoStop}%
\bibitem [{\citenamefont {Cea}\ \emph {et~al.}(2022)\citenamefont {Cea},
  \citenamefont {Pantaleón}, \citenamefont {Phong},\ and\ \citenamefont
  {Guinea}}]{cea_superconductivity_2022}%
  \BibitemOpen
  \bibfield  {author} {\bibinfo {author} {\bibfnamefont {T.}~\bibnamefont
  {Cea}}, \bibinfo {author} {\bibfnamefont {P.~A.}\ \bibnamefont {Pantaleón}},
  \bibinfo {author} {\bibfnamefont {V.~T.}\ \bibnamefont {Phong}},\ and\
  \bibinfo {author} {\bibfnamefont {F.}~\bibnamefont {Guinea}},\ }\href
  {https://doi.org/10.1103/PhysRevB.105.075432} {\bibfield  {journal} {\bibinfo
   {journal} {Physical Review B}\ }\textbf {\bibinfo {volume} {105}},\ \bibinfo
  {pages} {075432} (\bibinfo {year} {2022})}\BibitemShut {NoStop}%
\bibitem [{\citenamefont {Wagner}\ \emph {et~al.}(2023)\citenamefont {Wagner},
  \citenamefont {Kwan}, \citenamefont {Bultinck}, \citenamefont {Simon},\ and\
  \citenamefont {Parameswaran}}]{wagner_superconductivity_2023}%
  \BibitemOpen
  \bibfield  {author} {\bibinfo {author} {\bibfnamefont {G.}~\bibnamefont
  {Wagner}}, \bibinfo {author} {\bibfnamefont {Y.~H.}\ \bibnamefont {Kwan}},
  \bibinfo {author} {\bibfnamefont {N.}~\bibnamefont {Bultinck}}, \bibinfo
  {author} {\bibfnamefont {S.~H.}\ \bibnamefont {Simon}},\ and\ \bibinfo
  {author} {\bibfnamefont {S.~A.}\ \bibnamefont {Parameswaran}},\ }\href
  {https://doi.org/10.48550/arXiv.2302.00682} {\bibinfo {title}
  {Superconductivity from repulsive interactions in {Bernal}-stacked bilayer
  graphene}} (\bibinfo {year} {2023}),\ \bibinfo {note} {arXiv:2302.00682
  [cond-mat]}\BibitemShut {NoStop}%
\bibitem [{\citenamefont {Chatterjee}\ \emph {et~al.}(2022)\citenamefont
  {Chatterjee}, \citenamefont {Wang}, \citenamefont {Berg},\ and\ \citenamefont
  {Zaletel}}]{chatterjee_inter-valley_2022}%
  \BibitemOpen
  \bibfield  {author} {\bibinfo {author} {\bibfnamefont {S.}~\bibnamefont
  {Chatterjee}}, \bibinfo {author} {\bibfnamefont {T.}~\bibnamefont {Wang}},
  \bibinfo {author} {\bibfnamefont {E.}~\bibnamefont {Berg}},\ and\ \bibinfo
  {author} {\bibfnamefont {M.~P.}\ \bibnamefont {Zaletel}},\ }\href
  {https://doi.org/10.1038/s41467-022-33561-w} {\bibfield  {journal} {\bibinfo
  {journal} {Nature Communications}\ }\textbf {\bibinfo {volume} {13}},\
  \bibinfo {pages} {6013} (\bibinfo {year} {2022})}\BibitemShut {NoStop}%
\bibitem [{\citenamefont {Dong}\ and\ \citenamefont
  {Levitov}(2021)}]{dong_superconductivity_2021}%
  \BibitemOpen
  \bibfield  {author} {\bibinfo {author} {\bibfnamefont {Z.}~\bibnamefont
  {Dong}}\ and\ \bibinfo {author} {\bibfnamefont {L.}~\bibnamefont {Levitov}},\
  }\href {http://arxiv.org/abs/2109.01133} {\bibfield  {journal} {\bibinfo
  {journal} {arXiv:2109.01133 [cond-mat]}\ } (\bibinfo {year} {2021})},\
  \bibinfo {note} {arXiv: 2109.01133 [cond-mat]}\BibitemShut {NoStop}%
\bibitem [{\citenamefont {Dong}\ \emph {et~al.}(2023)\citenamefont {Dong},
  \citenamefont {Lee},\ and\ \citenamefont {Levitov}}]{dong_signatures_2023}%
  \BibitemOpen
  \bibfield  {author} {\bibinfo {author} {\bibfnamefont {Z.}~\bibnamefont
  {Dong}}, \bibinfo {author} {\bibfnamefont {P.~A.}\ \bibnamefont {Lee}},\ and\
  \bibinfo {author} {\bibfnamefont {L.~S.}\ \bibnamefont {Levitov}},\ }\href
  {https://doi.org/10.48550/arXiv.2304.09812} {\bibinfo {title} {Signatures of
  {Cooper} pair dynamics and quantum-critical superconductivity in tunable
  carrier bands}} (\bibinfo {year} {2023}),\ \bibinfo {note} {arXiv:2304.09812
  [cond-mat]}\BibitemShut {NoStop}%
\bibitem [{\citenamefont {Dong}\ \emph {et~al.}(2022)\citenamefont {Dong},
  \citenamefont {Chubukov},\ and\ \citenamefont
  {Levitov}}]{dong_spin-triplet_2022}%
  \BibitemOpen
  \bibfield  {author} {\bibinfo {author} {\bibfnamefont {Z.}~\bibnamefont
  {Dong}}, \bibinfo {author} {\bibfnamefont {A.~V.}\ \bibnamefont {Chubukov}},\
  and\ \bibinfo {author} {\bibfnamefont {L.}~\bibnamefont {Levitov}},\ }\href
  {http://arxiv.org/abs/2205.13353} {\bibinfo {title} {Spin-triplet
  superconductivity at the onset of isospin order in biased bilayer graphene}}
  (\bibinfo {year} {2022}),\ \bibinfo {note} {arXiv:2205.13353
  [cond-mat]}\BibitemShut {NoStop}%
\bibitem [{\citenamefont {Chou}\ \emph {et~al.}(2021)\citenamefont {Chou},
  \citenamefont {Wu}, \citenamefont {Sau},\ and\ \citenamefont
  {Sarma}}]{chou_acoustic-phonon-mediated_2021}%
  \BibitemOpen
  \bibfield  {author} {\bibinfo {author} {\bibfnamefont {Y.-Z.}\ \bibnamefont
  {Chou}}, \bibinfo {author} {\bibfnamefont {F.}~\bibnamefont {Wu}}, \bibinfo
  {author} {\bibfnamefont {J.~D.}\ \bibnamefont {Sau}},\ and\ \bibinfo {author}
  {\bibfnamefont {S.~D.}\ \bibnamefont {Sarma}},\ }\href@noop {} {\bibinfo
  {title} {Acoustic-phonon-mediated superconductivity in rhombohedral trilayer
  graphene}} (\bibinfo {year} {2021}),\ \bibinfo {note} {arXiv:2106.13231
  [cond-mat]}\BibitemShut {NoStop}%
\bibitem [{\citenamefont {Eisenstein}\ \emph {et~al.}(1992)\citenamefont
  {Eisenstein}, \citenamefont {Pfeiffer},\ and\ \citenamefont
  {West}}]{eisenstein_negative_1992}%
  \BibitemOpen
  \bibfield  {author} {\bibinfo {author} {\bibfnamefont {J.~P.}\ \bibnamefont
  {Eisenstein}}, \bibinfo {author} {\bibfnamefont {L.~N.}\ \bibnamefont
  {Pfeiffer}},\ and\ \bibinfo {author} {\bibfnamefont {K.~W.}\ \bibnamefont
  {West}},\ }\href {https://doi.org/10.1103/PhysRevLett.68.674} {\bibfield
  {journal} {\bibinfo  {journal} {Phys. Rev. Lett.}\ }\textbf {\bibinfo
  {volume} {68}},\ \bibinfo {pages} {674} (\bibinfo {year} {1992})}\BibitemShut
  {NoStop}%
\bibitem [{\citenamefont {Ashoori}\ \emph {et~al.}(1992)\citenamefont
  {Ashoori}, \citenamefont {Stormer}, \citenamefont {Weiner}, \citenamefont
  {Pfeiffer}, \citenamefont {Pearton}, \citenamefont {Baldwin},\ and\
  \citenamefont {West}}]{ashoori_single-electron_1992}%
  \BibitemOpen
  \bibfield  {author} {\bibinfo {author} {\bibfnamefont {R.~C.}\ \bibnamefont
  {Ashoori}}, \bibinfo {author} {\bibfnamefont {H.~L.}\ \bibnamefont
  {Stormer}}, \bibinfo {author} {\bibfnamefont {J.~S.}\ \bibnamefont {Weiner}},
  \bibinfo {author} {\bibfnamefont {L.~N.}\ \bibnamefont {Pfeiffer}}, \bibinfo
  {author} {\bibfnamefont {S.~J.}\ \bibnamefont {Pearton}}, \bibinfo {author}
  {\bibfnamefont {K.~W.}\ \bibnamefont {Baldwin}},\ and\ \bibinfo {author}
  {\bibfnamefont {K.~W.}\ \bibnamefont {West}},\ }\href
  {http://link.aps.org/doi/10.1103/PhysRevLett.68.3088} {\bibfield  {journal}
  {\bibinfo  {journal} {Phys. Rev. Lett.}\ }\textbf {\bibinfo {volume} {68}},\
  \bibinfo {pages} {3088} (\bibinfo {year} {1992})}\BibitemShut {NoStop}%
\bibitem [{\citenamefont {Das}\ and\ \citenamefont
  {Huang}(2023{\natexlab{a}})}]{das_unconventional_2023}%
  \BibitemOpen
  \bibfield  {author} {\bibinfo {author} {\bibfnamefont {M.}~\bibnamefont
  {Das}}\ and\ \bibinfo {author} {\bibfnamefont {C.}~\bibnamefont {Huang}},\
  }\href {http://arxiv.org/abs/2308.01996} {\bibinfo {title} {Unconventional
  {Metallic} {Magnetism}: {Non}-analyticity and {Sign}-changing {Behavior} of
  {Orbital} {Magnetization} in {ABC} {Trilayer} {Graphene}}} (\bibinfo {year}
  {2023}{\natexlab{a}}),\ \bibinfo {note} {arXiv:2308.01996
  [cond-mat]}\BibitemShut {NoStop}%
\bibitem [{\citenamefont {Coissard}\ \emph {et~al.}(2022)\citenamefont
  {Coissard}, \citenamefont {Wander}, \citenamefont {Vignaud}, \citenamefont
  {Grushin}, \citenamefont {Repellin}, \citenamefont {Watanabe}, \citenamefont
  {Taniguchi}, \citenamefont {Gay}, \citenamefont {Winkelmann}, \citenamefont
  {Courtois}, \citenamefont {Sellier},\ and\ \citenamefont
  {Sacépé}}]{coissard_imaging_2022}%
  \BibitemOpen
  \bibfield  {author} {\bibinfo {author} {\bibfnamefont {A.}~\bibnamefont
  {Coissard}}, \bibinfo {author} {\bibfnamefont {D.}~\bibnamefont {Wander}},
  \bibinfo {author} {\bibfnamefont {H.}~\bibnamefont {Vignaud}}, \bibinfo
  {author} {\bibfnamefont {A.~G.}\ \bibnamefont {Grushin}}, \bibinfo {author}
  {\bibfnamefont {C.}~\bibnamefont {Repellin}}, \bibinfo {author}
  {\bibfnamefont {K.}~\bibnamefont {Watanabe}}, \bibinfo {author}
  {\bibfnamefont {T.}~\bibnamefont {Taniguchi}}, \bibinfo {author}
  {\bibfnamefont {F.}~\bibnamefont {Gay}}, \bibinfo {author} {\bibfnamefont
  {C.~B.}\ \bibnamefont {Winkelmann}}, \bibinfo {author} {\bibfnamefont
  {H.}~\bibnamefont {Courtois}}, \bibinfo {author} {\bibfnamefont
  {H.}~\bibnamefont {Sellier}},\ and\ \bibinfo {author} {\bibfnamefont
  {B.}~\bibnamefont {Sacépé}},\ }\href
  {https://doi.org/10.1038/s41586-022-04513-7} {\bibfield  {journal} {\bibinfo
  {journal} {Nature}\ }\textbf {\bibinfo {volume} {605}},\ \bibinfo {pages}
  {51} (\bibinfo {year} {2022})}\BibitemShut {NoStop}%
\bibitem [{\citenamefont {Liu}\ \emph {et~al.}(2022)\citenamefont {Liu},
  \citenamefont {Farahi}, \citenamefont {Chiu}, \citenamefont {Papic},
  \citenamefont {Watanabe}, \citenamefont {Taniguchi}, \citenamefont
  {Zaletel},\ and\ \citenamefont {Yazdani}}]{liu_visualizing_2022}%
  \BibitemOpen
  \bibfield  {author} {\bibinfo {author} {\bibfnamefont {X.}~\bibnamefont
  {Liu}}, \bibinfo {author} {\bibfnamefont {G.}~\bibnamefont {Farahi}},
  \bibinfo {author} {\bibfnamefont {C.-L.}\ \bibnamefont {Chiu}}, \bibinfo
  {author} {\bibfnamefont {Z.}~\bibnamefont {Papic}}, \bibinfo {author}
  {\bibfnamefont {K.}~\bibnamefont {Watanabe}}, \bibinfo {author}
  {\bibfnamefont {T.}~\bibnamefont {Taniguchi}}, \bibinfo {author}
  {\bibfnamefont {M.~P.}\ \bibnamefont {Zaletel}},\ and\ \bibinfo {author}
  {\bibfnamefont {A.}~\bibnamefont {Yazdani}},\ }\href
  {https://doi.org/10.1126/science.abm3770} {\bibfield  {journal} {\bibinfo
  {journal} {Science}\ }\textbf {\bibinfo {volume} {375}},\ \bibinfo {pages}
  {321} (\bibinfo {year} {2022})}\BibitemShut {NoStop}%
\bibitem [{\citenamefont {Nuckolls}\ \emph {et~al.}(2023)\citenamefont
  {Nuckolls}, \citenamefont {Lee}, \citenamefont {Oh}, \citenamefont {Wong},
  \citenamefont {Soejima}, \citenamefont {Hong}, \citenamefont {Călugăru},
  \citenamefont {Herzog-Arbeitman}, \citenamefont {Bernevig}, \citenamefont
  {Watanabe}, \citenamefont {Taniguchi}, \citenamefont {Regnault},
  \citenamefont {Zaletel},\ and\ \citenamefont
  {Yazdani}}]{nuckolls_quantum_2023}%
  \BibitemOpen
  \bibfield  {author} {\bibinfo {author} {\bibfnamefont {K.~P.}\ \bibnamefont
  {Nuckolls}}, \bibinfo {author} {\bibfnamefont {R.~L.}\ \bibnamefont {Lee}},
  \bibinfo {author} {\bibfnamefont {M.}~\bibnamefont {Oh}}, \bibinfo {author}
  {\bibfnamefont {D.}~\bibnamefont {Wong}}, \bibinfo {author} {\bibfnamefont
  {T.}~\bibnamefont {Soejima}}, \bibinfo {author} {\bibfnamefont {J.~P.}\
  \bibnamefont {Hong}}, \bibinfo {author} {\bibfnamefont {D.}~\bibnamefont
  {Călugăru}}, \bibinfo {author} {\bibfnamefont {J.}~\bibnamefont
  {Herzog-Arbeitman}}, \bibinfo {author} {\bibfnamefont {B.~A.}\ \bibnamefont
  {Bernevig}}, \bibinfo {author} {\bibfnamefont {K.}~\bibnamefont {Watanabe}},
  \bibinfo {author} {\bibfnamefont {T.}~\bibnamefont {Taniguchi}}, \bibinfo
  {author} {\bibfnamefont {N.}~\bibnamefont {Regnault}}, \bibinfo {author}
  {\bibfnamefont {M.~P.}\ \bibnamefont {Zaletel}},\ and\ \bibinfo {author}
  {\bibfnamefont {A.}~\bibnamefont {Yazdani}},\ }\href
  {https://doi.org/10.1038/s41586-023-06226-x} {\bibfield  {journal} {\bibinfo
  {journal} {Nature}\ }\textbf {\bibinfo {volume} {620}},\ \bibinfo {pages}
  {525} (\bibinfo {year} {2023})}\BibitemShut {NoStop}%
\bibitem [{\citenamefont {Kim}\ \emph {et~al.}(2023)\citenamefont {Kim},
  \citenamefont {Choi}, \citenamefont {Lantagne-Hurtubise}, \citenamefont
  {Lewandowski}, \citenamefont {Thomson}, \citenamefont {Kong}, \citenamefont
  {Zhou}, \citenamefont {Baum}, \citenamefont {Zhang}, \citenamefont {Holleis},
  \citenamefont {Watanabe}, \citenamefont {Taniguchi}, \citenamefont {Young},
  \citenamefont {Alicea},\ and\ \citenamefont {Nadj-Perge}}]{kim_imaging_2023}%
  \BibitemOpen
  \bibfield  {author} {\bibinfo {author} {\bibfnamefont {H.}~\bibnamefont
  {Kim}}, \bibinfo {author} {\bibfnamefont {Y.}~\bibnamefont {Choi}}, \bibinfo
  {author} {\bibfnamefont {E.}~\bibnamefont {Lantagne-Hurtubise}}, \bibinfo
  {author} {\bibfnamefont {C.}~\bibnamefont {Lewandowski}}, \bibinfo {author}
  {\bibfnamefont {A.}~\bibnamefont {Thomson}}, \bibinfo {author} {\bibfnamefont
  {L.}~\bibnamefont {Kong}}, \bibinfo {author} {\bibfnamefont {H.}~\bibnamefont
  {Zhou}}, \bibinfo {author} {\bibfnamefont {E.}~\bibnamefont {Baum}}, \bibinfo
  {author} {\bibfnamefont {Y.}~\bibnamefont {Zhang}}, \bibinfo {author}
  {\bibfnamefont {L.}~\bibnamefont {Holleis}}, \bibinfo {author} {\bibfnamefont
  {K.}~\bibnamefont {Watanabe}}, \bibinfo {author} {\bibfnamefont
  {T.}~\bibnamefont {Taniguchi}}, \bibinfo {author} {\bibfnamefont {A.~F.}\
  \bibnamefont {Young}}, \bibinfo {author} {\bibfnamefont {J.}~\bibnamefont
  {Alicea}},\ and\ \bibinfo {author} {\bibfnamefont {S.}~\bibnamefont
  {Nadj-Perge}},\ }\href {https://doi.org/10.48550/arXiv.2304.10586} {\bibinfo
  {title} {Imaging inter-valley coherent order in magic-angle twisted trilayer
  graphene}} (\bibinfo {year} {2023}),\ \bibinfo {note} {arXiv:2304.10586
  [cond-mat]}\BibitemShut {NoStop}%
\bibitem [{\citenamefont {Zhang}\ \emph {et~al.}(2009)\citenamefont {Zhang},
  \citenamefont {Tang}, \citenamefont {Girit}, \citenamefont {Hao},
  \citenamefont {Martin}, \citenamefont {Zettl}, \citenamefont {Crommie},
  \citenamefont {Shen},\ and\ \citenamefont {Wang}}]{zhang_direct_2009}%
  \BibitemOpen
  \bibfield  {author} {\bibinfo {author} {\bibfnamefont {Y.}~\bibnamefont
  {Zhang}}, \bibinfo {author} {\bibfnamefont {T.-T.}\ \bibnamefont {Tang}},
  \bibinfo {author} {\bibfnamefont {C.}~\bibnamefont {Girit}}, \bibinfo
  {author} {\bibfnamefont {Z.}~\bibnamefont {Hao}}, \bibinfo {author}
  {\bibfnamefont {M.~C.}\ \bibnamefont {Martin}}, \bibinfo {author}
  {\bibfnamefont {A.}~\bibnamefont {Zettl}}, \bibinfo {author} {\bibfnamefont
  {M.~F.}\ \bibnamefont {Crommie}}, \bibinfo {author} {\bibfnamefont {Y.~R.}\
  \bibnamefont {Shen}},\ and\ \bibinfo {author} {\bibfnamefont
  {F.}~\bibnamefont {Wang}},\ }\href {http://dx.doi.org/10.1038/nature08105}
  {\bibfield  {journal} {\bibinfo  {journal} {Nature}\ }\textbf {\bibinfo
  {volume} {459}},\ \bibinfo {pages} {820} (\bibinfo {year}
  {2009})}\BibitemShut {NoStop}%
\bibitem [{\citenamefont {Ochoa}\ \emph {et~al.}(2012)\citenamefont {Ochoa},
  \citenamefont {Castro~Neto}, \citenamefont {Fal'ko},\ and\ \citenamefont
  {Guinea}}]{ochoa_spin-orbit_2012}%
  \BibitemOpen
  \bibfield  {author} {\bibinfo {author} {\bibfnamefont {H.}~\bibnamefont
  {Ochoa}}, \bibinfo {author} {\bibfnamefont {A.~H.}\ \bibnamefont
  {Castro~Neto}}, \bibinfo {author} {\bibfnamefont {V.~I.}\ \bibnamefont
  {Fal'ko}},\ and\ \bibinfo {author} {\bibfnamefont {F.}~\bibnamefont
  {Guinea}},\ }\href {https://doi.org/10.1103/PhysRevB.86.245411} {\bibfield
  {journal} {\bibinfo  {journal} {Physical Review B}\ }\textbf {\bibinfo
  {volume} {86}},\ \bibinfo {pages} {245411} (\bibinfo {year}
  {2012})}\BibitemShut {NoStop}%
\bibitem [{\citenamefont {Sichau}\ \emph {et~al.}(2019)\citenamefont {Sichau},
  \citenamefont {Prada}, \citenamefont {Anlauf}, \citenamefont {Lyon},
  \citenamefont {Bosnjak}, \citenamefont {Tiemann},\ and\ \citenamefont
  {Blick}}]{sichau_resonance_2019}%
  \BibitemOpen
  \bibfield  {author} {\bibinfo {author} {\bibfnamefont {J.}~\bibnamefont
  {Sichau}}, \bibinfo {author} {\bibfnamefont {M.}~\bibnamefont {Prada}},
  \bibinfo {author} {\bibfnamefont {T.}~\bibnamefont {Anlauf}}, \bibinfo
  {author} {\bibfnamefont {T.}~\bibnamefont {Lyon}}, \bibinfo {author}
  {\bibfnamefont {B.}~\bibnamefont {Bosnjak}}, \bibinfo {author} {\bibfnamefont
  {L.}~\bibnamefont {Tiemann}},\ and\ \bibinfo {author} {\bibfnamefont
  {R.}~\bibnamefont {Blick}},\ }\href
  {https://doi.org/10.1103/PhysRevLett.122.046403} {\bibfield  {journal}
  {\bibinfo  {journal} {Physical Review Letters}\ }\textbf {\bibinfo {volume}
  {122}},\ \bibinfo {pages} {046403} (\bibinfo {year} {2019})}\BibitemShut
  {NoStop}%
\bibitem [{\citenamefont {Banszerus}\ \emph {et~al.}(2020)\citenamefont
  {Banszerus}, \citenamefont {Frohn}, \citenamefont {Fabian}, \citenamefont
  {Somanchi}, \citenamefont {Epping}, \citenamefont {Müller}, \citenamefont
  {Neumaier}, \citenamefont {Watanabe}, \citenamefont {Taniguchi},
  \citenamefont {Libisch}, \citenamefont {Beschoten}, \citenamefont {Hassler},\
  and\ \citenamefont {Stampfer}}]{banszerus_observation_2020}%
  \BibitemOpen
  \bibfield  {author} {\bibinfo {author} {\bibfnamefont {L.}~\bibnamefont
  {Banszerus}}, \bibinfo {author} {\bibfnamefont {B.}~\bibnamefont {Frohn}},
  \bibinfo {author} {\bibfnamefont {T.}~\bibnamefont {Fabian}}, \bibinfo
  {author} {\bibfnamefont {S.}~\bibnamefont {Somanchi}}, \bibinfo {author}
  {\bibfnamefont {A.}~\bibnamefont {Epping}}, \bibinfo {author} {\bibfnamefont
  {M.}~\bibnamefont {Müller}}, \bibinfo {author} {\bibfnamefont
  {D.}~\bibnamefont {Neumaier}}, \bibinfo {author} {\bibfnamefont
  {K.}~\bibnamefont {Watanabe}}, \bibinfo {author} {\bibfnamefont
  {T.}~\bibnamefont {Taniguchi}}, \bibinfo {author} {\bibfnamefont
  {F.}~\bibnamefont {Libisch}}, \bibinfo {author} {\bibfnamefont
  {B.}~\bibnamefont {Beschoten}}, \bibinfo {author} {\bibfnamefont
  {F.}~\bibnamefont {Hassler}},\ and\ \bibinfo {author} {\bibfnamefont
  {C.}~\bibnamefont {Stampfer}},\ }\href
  {https://doi.org/10.1103/PhysRevLett.124.177701} {\bibfield  {journal}
  {\bibinfo  {journal} {Physical Review Letters}\ }\textbf {\bibinfo {volume}
  {124}},\ \bibinfo {pages} {177701} (\bibinfo {year} {2020})}\BibitemShut
  {NoStop}%
\bibitem [{\citenamefont {Kurzmann}\ \emph {et~al.}(2021)\citenamefont
  {Kurzmann}, \citenamefont {Kleeorin}, \citenamefont {Tong}, \citenamefont
  {Garreis}, \citenamefont {Knothe}, \citenamefont {Eich}, \citenamefont
  {Mittag}, \citenamefont {Gold}, \citenamefont {de~Vries}, \citenamefont
  {Watanabe}, \citenamefont {Taniguchi}, \citenamefont {Fal’ko},
  \citenamefont {Meir}, \citenamefont {Ihn},\ and\ \citenamefont
  {Ensslin}}]{kurzmann_kondo_2021}%
  \BibitemOpen
  \bibfield  {author} {\bibinfo {author} {\bibfnamefont {A.}~\bibnamefont
  {Kurzmann}}, \bibinfo {author} {\bibfnamefont {Y.}~\bibnamefont {Kleeorin}},
  \bibinfo {author} {\bibfnamefont {C.}~\bibnamefont {Tong}}, \bibinfo {author}
  {\bibfnamefont {R.}~\bibnamefont {Garreis}}, \bibinfo {author} {\bibfnamefont
  {A.}~\bibnamefont {Knothe}}, \bibinfo {author} {\bibfnamefont
  {M.}~\bibnamefont {Eich}}, \bibinfo {author} {\bibfnamefont {C.}~\bibnamefont
  {Mittag}}, \bibinfo {author} {\bibfnamefont {C.}~\bibnamefont {Gold}},
  \bibinfo {author} {\bibfnamefont {F.~K.}\ \bibnamefont {de~Vries}}, \bibinfo
  {author} {\bibfnamefont {K.}~\bibnamefont {Watanabe}}, \bibinfo {author}
  {\bibfnamefont {T.}~\bibnamefont {Taniguchi}}, \bibinfo {author}
  {\bibfnamefont {V.}~\bibnamefont {Fal’ko}}, \bibinfo {author}
  {\bibfnamefont {Y.}~\bibnamefont {Meir}}, \bibinfo {author} {\bibfnamefont
  {T.}~\bibnamefont {Ihn}},\ and\ \bibinfo {author} {\bibfnamefont
  {K.}~\bibnamefont {Ensslin}},\ }\href
  {https://doi.org/10.1038/s41467-021-26149-3} {\bibfield  {journal} {\bibinfo
  {journal} {Nature Communications}\ }\textbf {\bibinfo {volume} {12}},\
  \bibinfo {pages} {6004} (\bibinfo {year} {2021})}\BibitemShut {NoStop}%
\bibitem [{\citenamefont {Lee}\ \emph {et~al.}(2019)\citenamefont {Lee},
  \citenamefont {Khalaf}, \citenamefont {Liu}, \citenamefont {Liu},
  \citenamefont {Hao}, \citenamefont {Kim},\ and\ \citenamefont
  {Vishwanath}}]{lee_theory_2019}%
  \BibitemOpen
  \bibfield  {author} {\bibinfo {author} {\bibfnamefont {J.~Y.}\ \bibnamefont
  {Lee}}, \bibinfo {author} {\bibfnamefont {E.}~\bibnamefont {Khalaf}},
  \bibinfo {author} {\bibfnamefont {S.}~\bibnamefont {Liu}}, \bibinfo {author}
  {\bibfnamefont {X.}~\bibnamefont {Liu}}, \bibinfo {author} {\bibfnamefont
  {Z.}~\bibnamefont {Hao}}, \bibinfo {author} {\bibfnamefont {P.}~\bibnamefont
  {Kim}},\ and\ \bibinfo {author} {\bibfnamefont {A.}~\bibnamefont
  {Vishwanath}},\ }\href {https://doi.org/10.1038/s41467-019-12981-1}
  {\bibfield  {journal} {\bibinfo  {journal} {Nature Communications}\ }\textbf
  {\bibinfo {volume} {10}},\ \bibinfo {pages} {5333} (\bibinfo {year}
  {2019})},\ \bibinfo {note} {number: 1 Publisher: Nature Publishing
  Group}\BibitemShut {NoStop}%
\bibitem [{\citenamefont {Cao}\ \emph {et~al.}(2018)\citenamefont {Cao},
  \citenamefont {Fatemi}, \citenamefont {Fang}, \citenamefont {Watanabe},
  \citenamefont {Taniguchi}, \citenamefont {Kaxiras},\ and\ \citenamefont
  {Jarillo-Herrero}}]{cao_unconventional_2018}%
  \BibitemOpen
  \bibfield  {author} {\bibinfo {author} {\bibfnamefont {Y.}~\bibnamefont
  {Cao}}, \bibinfo {author} {\bibfnamefont {V.}~\bibnamefont {Fatemi}},
  \bibinfo {author} {\bibfnamefont {S.}~\bibnamefont {Fang}}, \bibinfo {author}
  {\bibfnamefont {K.}~\bibnamefont {Watanabe}}, \bibinfo {author}
  {\bibfnamefont {T.}~\bibnamefont {Taniguchi}}, \bibinfo {author}
  {\bibfnamefont {E.}~\bibnamefont {Kaxiras}},\ and\ \bibinfo {author}
  {\bibfnamefont {P.}~\bibnamefont {Jarillo-Herrero}},\ }\href
  {https://doi.org/10.1038/nature26160} {\bibfield  {journal} {\bibinfo
  {journal} {Nature}\ }\textbf {\bibinfo {volume} {556}},\ \bibinfo {pages}
  {43} (\bibinfo {year} {2018})}\BibitemShut {NoStop}%
\bibitem [{\citenamefont {Park}\ \emph {et~al.}(2021)\citenamefont {Park},
  \citenamefont {Cao}, \citenamefont {Watanabe}, \citenamefont {Taniguchi},\
  and\ \citenamefont {Jarillo-Herrero}}]{park_tunable_2021}%
  \BibitemOpen
  \bibfield  {author} {\bibinfo {author} {\bibfnamefont {J.~M.}\ \bibnamefont
  {Park}}, \bibinfo {author} {\bibfnamefont {Y.}~\bibnamefont {Cao}}, \bibinfo
  {author} {\bibfnamefont {K.}~\bibnamefont {Watanabe}}, \bibinfo {author}
  {\bibfnamefont {T.}~\bibnamefont {Taniguchi}},\ and\ \bibinfo {author}
  {\bibfnamefont {P.}~\bibnamefont {Jarillo-Herrero}},\ }\href
  {https://doi.org/10.1038/s41586-021-03192-0} {\bibfield  {journal} {\bibinfo
  {journal} {Nature}\ }\textbf {\bibinfo {volume} {590}},\ \bibinfo {pages}
  {249} (\bibinfo {year} {2021})}\BibitemShut {NoStop}%
\bibitem [{\citenamefont {Hao}\ \emph {et~al.}(2021)\citenamefont {Hao},
  \citenamefont {Zimmerman}, \citenamefont {Ledwith}, \citenamefont {Khalaf},
  \citenamefont {Najafabadi}, \citenamefont {Watanabe}, \citenamefont
  {Taniguchi}, \citenamefont {Vishwanath},\ and\ \citenamefont
  {Kim}}]{hao_electric_2021}%
  \BibitemOpen
  \bibfield  {author} {\bibinfo {author} {\bibfnamefont {Z.}~\bibnamefont
  {Hao}}, \bibinfo {author} {\bibfnamefont {A.~M.}\ \bibnamefont {Zimmerman}},
  \bibinfo {author} {\bibfnamefont {P.}~\bibnamefont {Ledwith}}, \bibinfo
  {author} {\bibfnamefont {E.}~\bibnamefont {Khalaf}}, \bibinfo {author}
  {\bibfnamefont {D.~H.}\ \bibnamefont {Najafabadi}}, \bibinfo {author}
  {\bibfnamefont {K.}~\bibnamefont {Watanabe}}, \bibinfo {author}
  {\bibfnamefont {T.}~\bibnamefont {Taniguchi}}, \bibinfo {author}
  {\bibfnamefont {A.}~\bibnamefont {Vishwanath}},\ and\ \bibinfo {author}
  {\bibfnamefont {P.}~\bibnamefont {Kim}},\ }\href
  {https://doi.org/10.1126/science.abg0399} {\bibfield  {journal} {\bibinfo
  {journal} {Science}\ }\textbf {\bibinfo {volume} {371}},\ \bibinfo {pages}
  {1133} (\bibinfo {year} {2021})}\BibitemShut {NoStop}%
\bibitem [{\citenamefont {Arora}\ \emph {et~al.}(2020)\citenamefont {Arora},
  \citenamefont {Polski}, \citenamefont {Zhang}, \citenamefont {Thomson},
  \citenamefont {Choi}, \citenamefont {Kim}, \citenamefont {Lin}, \citenamefont
  {Wilson}, \citenamefont {Xu}, \citenamefont {Chu}, \citenamefont {Watanabe},
  \citenamefont {Taniguchi}, \citenamefont {Alicea},\ and\ \citenamefont
  {Nadj-Perge}}]{arora_superconductivity_2020}%
  \BibitemOpen
  \bibfield  {author} {\bibinfo {author} {\bibfnamefont {H.~S.}\ \bibnamefont
  {Arora}}, \bibinfo {author} {\bibfnamefont {R.}~\bibnamefont {Polski}},
  \bibinfo {author} {\bibfnamefont {Y.}~\bibnamefont {Zhang}}, \bibinfo
  {author} {\bibfnamefont {A.}~\bibnamefont {Thomson}}, \bibinfo {author}
  {\bibfnamefont {Y.}~\bibnamefont {Choi}}, \bibinfo {author} {\bibfnamefont
  {H.}~\bibnamefont {Kim}}, \bibinfo {author} {\bibfnamefont {Z.}~\bibnamefont
  {Lin}}, \bibinfo {author} {\bibfnamefont {I.~Z.}\ \bibnamefont {Wilson}},
  \bibinfo {author} {\bibfnamefont {X.}~\bibnamefont {Xu}}, \bibinfo {author}
  {\bibfnamefont {J.-H.}\ \bibnamefont {Chu}}, \bibinfo {author} {\bibfnamefont
  {K.}~\bibnamefont {Watanabe}}, \bibinfo {author} {\bibfnamefont
  {T.}~\bibnamefont {Taniguchi}}, \bibinfo {author} {\bibfnamefont
  {J.}~\bibnamefont {Alicea}},\ and\ \bibinfo {author} {\bibfnamefont
  {S.}~\bibnamefont {Nadj-Perge}},\ }\href
  {https://doi.org/10.1038/s41586-020-2473-8} {\bibfield  {journal} {\bibinfo
  {journal} {Nature}\ }\textbf {\bibinfo {volume} {583}},\ \bibinfo {pages}
  {379} (\bibinfo {year} {2020})}\BibitemShut {NoStop}%
\bibitem [{\citenamefont {Zhang}\ \emph {et~al.}(2022)\citenamefont {Zhang},
  \citenamefont {Polski}, \citenamefont {Lewandowski}, \citenamefont {Thomson},
  \citenamefont {Peng}, \citenamefont {Choi}, \citenamefont {Kim},
  \citenamefont {Watanabe}, \citenamefont {Taniguchi}, \citenamefont {Alicea},
  \citenamefont {von Oppen}, \citenamefont {Refael},\ and\ \citenamefont
  {Nadj-Perge}}]{zhang_promotion_2022}%
  \BibitemOpen
  \bibfield  {author} {\bibinfo {author} {\bibfnamefont {Y.}~\bibnamefont
  {Zhang}}, \bibinfo {author} {\bibfnamefont {R.}~\bibnamefont {Polski}},
  \bibinfo {author} {\bibfnamefont {C.}~\bibnamefont {Lewandowski}}, \bibinfo
  {author} {\bibfnamefont {A.}~\bibnamefont {Thomson}}, \bibinfo {author}
  {\bibfnamefont {Y.}~\bibnamefont {Peng}}, \bibinfo {author} {\bibfnamefont
  {Y.}~\bibnamefont {Choi}}, \bibinfo {author} {\bibfnamefont {H.}~\bibnamefont
  {Kim}}, \bibinfo {author} {\bibfnamefont {K.}~\bibnamefont {Watanabe}},
  \bibinfo {author} {\bibfnamefont {T.}~\bibnamefont {Taniguchi}}, \bibinfo
  {author} {\bibfnamefont {J.}~\bibnamefont {Alicea}}, \bibinfo {author}
  {\bibfnamefont {F.}~\bibnamefont {von Oppen}}, \bibinfo {author}
  {\bibfnamefont {G.}~\bibnamefont {Refael}},\ and\ \bibinfo {author}
  {\bibfnamefont {S.}~\bibnamefont {Nadj-Perge}},\ }\href
  {https://doi.org/10.1126/science.abn8585} {\bibfield  {journal} {\bibinfo
  {journal} {Science}\ }\textbf {\bibinfo {volume} {377}},\ \bibinfo {pages}
  {1538} (\bibinfo {year} {2022})}\BibitemShut {NoStop}%
\bibitem [{\citenamefont {Park}\ \emph {et~al.}(2022)\citenamefont {Park},
  \citenamefont {Cao}, \citenamefont {Xia}, \citenamefont {Sun}, \citenamefont
  {Watanabe}, \citenamefont {Taniguchi},\ and\ \citenamefont
  {Jarillo-Herrero}}]{park_robust_2022}%
  \BibitemOpen
  \bibfield  {author} {\bibinfo {author} {\bibfnamefont {J.~M.}\ \bibnamefont
  {Park}}, \bibinfo {author} {\bibfnamefont {Y.}~\bibnamefont {Cao}}, \bibinfo
  {author} {\bibfnamefont {L.-Q.}\ \bibnamefont {Xia}}, \bibinfo {author}
  {\bibfnamefont {S.}~\bibnamefont {Sun}}, \bibinfo {author} {\bibfnamefont
  {K.}~\bibnamefont {Watanabe}}, \bibinfo {author} {\bibfnamefont
  {T.}~\bibnamefont {Taniguchi}},\ and\ \bibinfo {author} {\bibfnamefont
  {P.}~\bibnamefont {Jarillo-Herrero}},\ }\href
  {https://doi.org/10.1038/s41563-022-01287-1} {\bibfield  {journal} {\bibinfo
  {journal} {Nature Materials}\ }\textbf {\bibinfo {volume} {21}},\ \bibinfo
  {pages} {877} (\bibinfo {year} {2022})}\BibitemShut {NoStop}%
\bibitem [{\citenamefont {Koh}\ \emph {et~al.}(2023)\citenamefont {Koh},
  \citenamefont {Alicea},\ and\ \citenamefont
  {Lantagne-Hurtubise}}]{koh_correlated_2023}%
  \BibitemOpen
  \bibfield  {author} {\bibinfo {author} {\bibfnamefont {J.~M.}\ \bibnamefont
  {Koh}}, \bibinfo {author} {\bibfnamefont {J.}~\bibnamefont {Alicea}},\ and\
  \bibinfo {author} {\bibfnamefont {E.}~\bibnamefont {Lantagne-Hurtubise}},\
  }\href {https://doi.org/10.48550/arXiv.2306.12486} {\bibinfo {title}
  {Correlated {Phases} in {Spin}-{Orbit}-{Coupled} {Rhombohedral} {Trilayer}
  {Graphene}}} (\bibinfo {year} {2023}),\ \bibinfo {note} {arXiv:2306.12486
  [cond-mat]}\BibitemShut {NoStop}%
\bibitem [{\citenamefont {Zhumagulov}\ \emph {et~al.}(2023)\citenamefont
  {Zhumagulov}, \citenamefont {Kochan},\ and\ \citenamefont
  {Fabian}}]{zhumagulov_emergent_2023}%
  \BibitemOpen
  \bibfield  {author} {\bibinfo {author} {\bibfnamefont {Y.}~\bibnamefont
  {Zhumagulov}}, \bibinfo {author} {\bibfnamefont {D.}~\bibnamefont {Kochan}},\
  and\ \bibinfo {author} {\bibfnamefont {J.}~\bibnamefont {Fabian}},\
  }\href@noop {} {\bibinfo {title} {Emergent correlated phases in rhombohedral
  trilayer graphene induced by proximity spin-orbit and exchange coupling}}
  (\bibinfo {year} {2023}),\ \bibinfo {note} {arXiv:2305.14277
  [condmat]}\BibitemShut {NoStop}%
\bibitem [{\citenamefont {Sharpe}\ \emph {et~al.}(2021)\citenamefont {Sharpe},
  \citenamefont {Fox}, \citenamefont {Barnard}, \citenamefont {Finney},
  \citenamefont {Watanabe}, \citenamefont {Taniguchi}, \citenamefont
  {Kastner},\ and\ \citenamefont {Goldhaber-Gordon}}]{sharpe_evidence_2021}%
  \BibitemOpen
  \bibfield  {author} {\bibinfo {author} {\bibfnamefont {A.~L.}\ \bibnamefont
  {Sharpe}}, \bibinfo {author} {\bibfnamefont {E.~J.}\ \bibnamefont {Fox}},
  \bibinfo {author} {\bibfnamefont {A.~W.}\ \bibnamefont {Barnard}}, \bibinfo
  {author} {\bibfnamefont {J.}~\bibnamefont {Finney}}, \bibinfo {author}
  {\bibfnamefont {K.}~\bibnamefont {Watanabe}}, \bibinfo {author}
  {\bibfnamefont {T.}~\bibnamefont {Taniguchi}}, \bibinfo {author}
  {\bibfnamefont {M.~A.}\ \bibnamefont {Kastner}},\ and\ \bibinfo {author}
  {\bibfnamefont {D.}~\bibnamefont {Goldhaber-Gordon}},\ }\href
  {https://doi.org/10.1021/acs.nanolett.1c00696} {\bibfield  {journal}
  {\bibinfo  {journal} {Nano Letters}\ }\textbf {\bibinfo {volume} {21}},\
  \bibinfo {pages} {4299} (\bibinfo {year} {2021})}\BibitemShut {NoStop}%
\bibitem [{\citenamefont {Kuiri}\ \emph {et~al.}(2022)\citenamefont {Kuiri},
  \citenamefont {Coleman}, \citenamefont {Gao}, \citenamefont {Vishnuradhan},
  \citenamefont {Watanabe}, \citenamefont {Taniguchi}, \citenamefont {Zhu},
  \citenamefont {MacDonald},\ and\ \citenamefont
  {Folk}}]{kuiri_spontaneous_2022}%
  \BibitemOpen
  \bibfield  {author} {\bibinfo {author} {\bibfnamefont {M.}~\bibnamefont
  {Kuiri}}, \bibinfo {author} {\bibfnamefont {C.}~\bibnamefont {Coleman}},
  \bibinfo {author} {\bibfnamefont {Z.}~\bibnamefont {Gao}}, \bibinfo {author}
  {\bibfnamefont {A.}~\bibnamefont {Vishnuradhan}}, \bibinfo {author}
  {\bibfnamefont {K.}~\bibnamefont {Watanabe}}, \bibinfo {author}
  {\bibfnamefont {T.}~\bibnamefont {Taniguchi}}, \bibinfo {author}
  {\bibfnamefont {J.}~\bibnamefont {Zhu}}, \bibinfo {author} {\bibfnamefont
  {A.~H.}\ \bibnamefont {MacDonald}},\ and\ \bibinfo {author} {\bibfnamefont
  {J.}~\bibnamefont {Folk}},\ }\href
  {https://doi.org/10.1038/s41467-022-34192-x} {\bibfield  {journal} {\bibinfo
  {journal} {Nature Communications}\ }\textbf {\bibinfo {volume} {13}},\
  \bibinfo {pages} {6468} (\bibinfo {year} {2022})}\BibitemShut {NoStop}%
\bibitem [{\citenamefont {Huang}\ \emph {et~al.}(2023)\citenamefont {Huang},
  \citenamefont {Wolf}, \citenamefont {Qin}, \citenamefont {Wei}, \citenamefont
  {Blinov},\ and\ \citenamefont {MacDonald}}]{huang_spin_2023}%
  \BibitemOpen
  \bibfield  {author} {\bibinfo {author} {\bibfnamefont {C.}~\bibnamefont
  {Huang}}, \bibinfo {author} {\bibfnamefont {T.~M.~R.}\ \bibnamefont {Wolf}},
  \bibinfo {author} {\bibfnamefont {W.}~\bibnamefont {Qin}}, \bibinfo {author}
  {\bibfnamefont {N.}~\bibnamefont {Wei}}, \bibinfo {author} {\bibfnamefont
  {I.~V.}\ \bibnamefont {Blinov}},\ and\ \bibinfo {author} {\bibfnamefont
  {A.~H.}\ \bibnamefont {MacDonald}},\ }\href
  {https://doi.org/10.1103/PhysRevB.107.L121405} {\bibfield  {journal}
  {\bibinfo  {journal} {Physical Review B}\ }\textbf {\bibinfo {volume}
  {107}},\ \bibinfo {pages} {L121405} (\bibinfo {year} {2023})}\BibitemShut
  {NoStop}%
\bibitem [{\citenamefont {Cao}\ \emph {et~al.}(2021)\citenamefont {Cao},
  \citenamefont {Park}, \citenamefont {Watanabe}, \citenamefont {Taniguchi},\
  and\ \citenamefont {Jarillo-Herrero}}]{cao_pauli-limit_2021}%
  \BibitemOpen
  \bibfield  {author} {\bibinfo {author} {\bibfnamefont {Y.}~\bibnamefont
  {Cao}}, \bibinfo {author} {\bibfnamefont {J.~M.}\ \bibnamefont {Park}},
  \bibinfo {author} {\bibfnamefont {K.}~\bibnamefont {Watanabe}}, \bibinfo
  {author} {\bibfnamefont {T.}~\bibnamefont {Taniguchi}},\ and\ \bibinfo
  {author} {\bibfnamefont {P.}~\bibnamefont {Jarillo-Herrero}},\ }\href
  {https://doi.org/10.1038/s41586-021-03685-y} {\bibfield  {journal} {\bibinfo
  {journal} {Nature}\ }\textbf {\bibinfo {volume} {595}},\ \bibinfo {pages}
  {526} (\bibinfo {year} {2021})}\BibitemShut {NoStop}%
\bibitem [{\citenamefont {Das}\ and\ \citenamefont
  {Huang}(2023{\natexlab{b}})}]{das_quarter-metal_2023}%
  \BibitemOpen
  \bibfield  {author} {\bibinfo {author} {\bibfnamefont {M.}~\bibnamefont
  {Das}}\ and\ \bibinfo {author} {\bibfnamefont {C.}~\bibnamefont {Huang}},\
  }\href {http://arxiv.org/abs/2310.10759} {\bibinfo {title} {Quarter-{Metal}
  {Phases} in {Multilayer} {Graphene}: {Ising}-{XY} and {Annular} {Lifshitz}
  {Transitions}}} (\bibinfo {year} {2023}{\natexlab{b}}),\ \bibinfo {note}
  {arXiv:2310.10759 [cond-mat]}\BibitemShut {NoStop}%
\bibitem [{\citenamefont {Anahory}\ \emph {et~al.}(2020)\citenamefont
  {Anahory}, \citenamefont {Naren}, \citenamefont {Lachman}, \citenamefont
  {Sinai}, \citenamefont {Uri}, \citenamefont {Embon}, \citenamefont {Yaakobi},
  \citenamefont {Myasoedov}, \citenamefont {Huber}, \citenamefont {Klajn},\
  and\ \citenamefont {Zeldov}}]{anahory_squid--tip_2020}%
  \BibitemOpen
  \bibfield  {author} {\bibinfo {author} {\bibfnamefont {Y.}~\bibnamefont
  {Anahory}}, \bibinfo {author} {\bibfnamefont {H.~R.}\ \bibnamefont {Naren}},
  \bibinfo {author} {\bibfnamefont {E.~O.}\ \bibnamefont {Lachman}}, \bibinfo
  {author} {\bibfnamefont {S.~B.}\ \bibnamefont {Sinai}}, \bibinfo {author}
  {\bibfnamefont {A.}~\bibnamefont {Uri}}, \bibinfo {author} {\bibfnamefont
  {L.}~\bibnamefont {Embon}}, \bibinfo {author} {\bibfnamefont
  {E.}~\bibnamefont {Yaakobi}}, \bibinfo {author} {\bibfnamefont
  {Y.}~\bibnamefont {Myasoedov}}, \bibinfo {author} {\bibfnamefont {M.~E.}\
  \bibnamefont {Huber}}, \bibinfo {author} {\bibfnamefont {R.}~\bibnamefont
  {Klajn}},\ and\ \bibinfo {author} {\bibfnamefont {E.}~\bibnamefont
  {Zeldov}},\ }\href {https://doi.org/10.1039/C9NR08578E} {\bibfield  {journal}
  {\bibinfo  {journal} {Nanoscale}\ }\textbf {\bibinfo {volume} {12}},\
  \bibinfo {pages} {3174} (\bibinfo {year} {2020})}\BibitemShut {NoStop}%
\bibitem [{\citenamefont {Zhang}\ \emph {et~al.}(2010)\citenamefont {Zhang},
  \citenamefont {Sahu}, \citenamefont {Min},\ and\ \citenamefont
  {MacDonald}}]{zhang_band_2010}%
  \BibitemOpen
  \bibfield  {author} {\bibinfo {author} {\bibfnamefont {F.}~\bibnamefont
  {Zhang}}, \bibinfo {author} {\bibfnamefont {B.}~\bibnamefont {Sahu}},
  \bibinfo {author} {\bibfnamefont {H.}~\bibnamefont {Min}},\ and\ \bibinfo
  {author} {\bibfnamefont {A.~H.}\ \bibnamefont {MacDonald}},\ }\href
  {https://doi.org/10.1103/PhysRevB.82.035409} {\bibfield  {journal} {\bibinfo
  {journal} {Physical Review B}\ }\textbf {\bibinfo {volume} {82}},\ \bibinfo
  {pages} {035409} (\bibinfo {year} {2010})}\BibitemShut {NoStop}%
\bibitem [{\citenamefont {Kudin}\ \emph {et~al.}(2002)\citenamefont {Kudin},
  \citenamefont {Scuseria},\ and\ \citenamefont
  {Cancès}}]{kudin_black-box_2002}%
  \BibitemOpen
  \bibfield  {author} {\bibinfo {author} {\bibfnamefont {K.~N.}\ \bibnamefont
  {Kudin}}, \bibinfo {author} {\bibfnamefont {G.~E.}\ \bibnamefont
  {Scuseria}},\ and\ \bibinfo {author} {\bibfnamefont {E.}~\bibnamefont
  {Cancès}},\ }\href {https://doi.org/10.1063/1.1470195} {\bibfield  {journal}
  {\bibinfo  {journal} {The Journal of Chemical Physics}\ }\textbf {\bibinfo
  {volume} {116}},\ \bibinfo {pages} {8255} (\bibinfo {year}
  {2002})}\BibitemShut {NoStop}%
\end{thebibliography}%

\clearpage
\newpage
\pagebreak

\onecolumngrid

\begin{center}
\textbf{\large Extended Data Figures}\\[5pt]
\end{center}

\begin{figure}[ht]
    \centering
    \includegraphics[width=183mm]{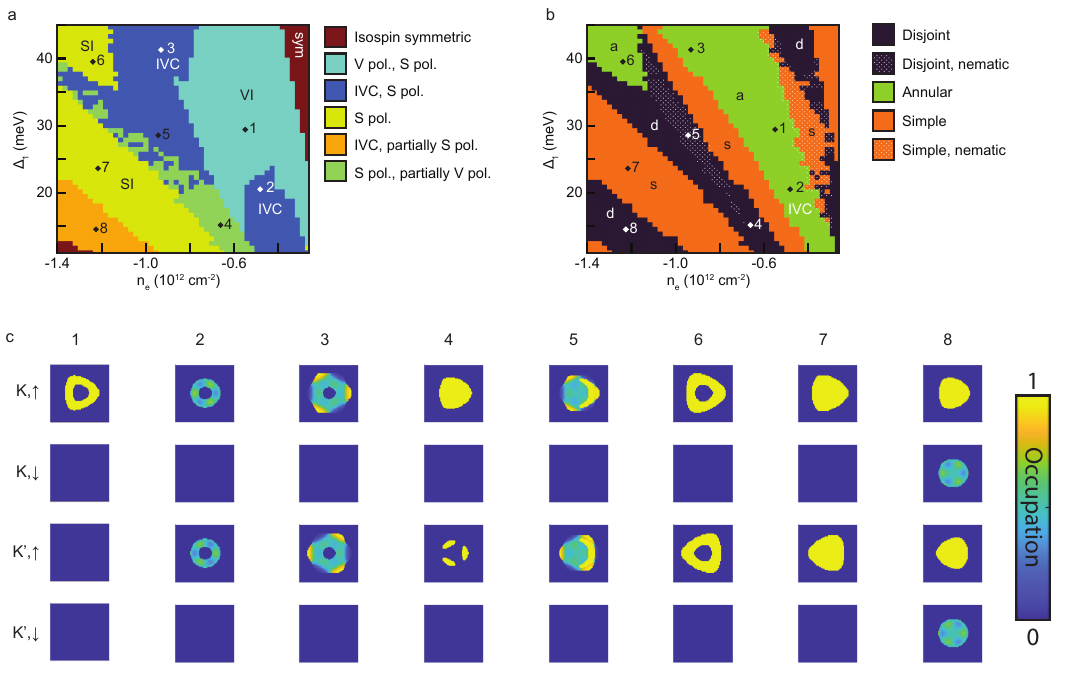}
    \caption{
    \textbf{Hartree Fock phase diagram for hole doping.} 
    \textbf{(A)} Phase diagram without SOC, showing the ground state isospin polarization as a function of $n_e$ and the interlayer potential $\Delta_1$. 
    \textbf{(B)} Phase diagram of Fermi surface topology from the same calculation. Regions that break the C$_3$ symmetry (i.e., nematics) are indicated by white dots. ``Disjoint'' refers to any case where there are multiple Fermi surfaces with different areas, excluding annuli.   
    \textbf{(C)} Flavor occupation of the Hartree-Fock ground state at the points labeled in panels A, B.  The color indicates occupation expectation values. The isospin order, Fermi surface topology, and nematicity in these eight points is summarized in the table below. Note that in point no. 5 there are two distinct Fermi surfaces with different areas, and hence the FS topology is ``disjoint'' in our scheme.  
\label{fig:newHF}
}
    \begin{tabular}{|c|c|c|c|}
 \hline 
 Point  &Isospin order&Topology&Nematic\\ \hline
1& VI, SI &Annular &\xmark\ \\ \hline
2& SI, IVC & Annular& \xmark \\ \hline
3& SI, IVC & Annular & \xmark \\ \hline
4& VI, SI  & Disjoint & \xmark \\ \hline
5& SI, IVC& Disjoint & \cmark \\\hline
6&  SI & Annular&  \xmark\\ \hline
7& SI  & Simple &  \xmark \\\hline
8& SI, IVC & Disjoint& \xmark \\ \hline
    \end{tabular}
\end{figure}

\begin{figure}[ht]
    \centering
    \includegraphics[width=183mm]{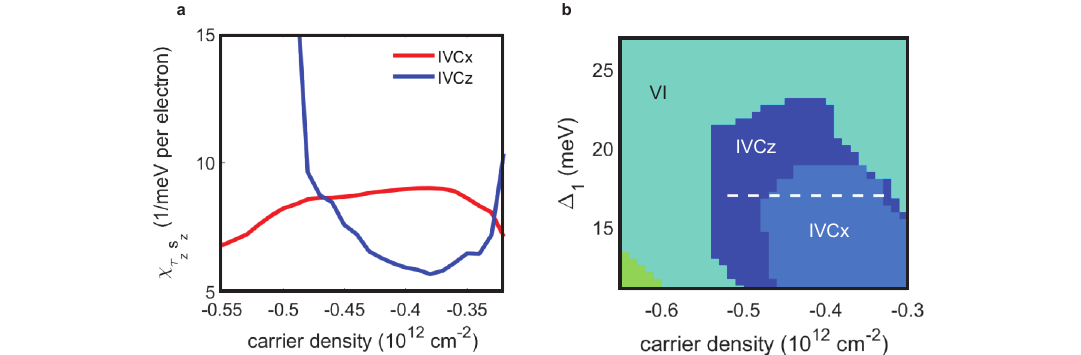}
    \caption{
    \textbf{Effects of spin-orbit coupling in the IVC phase from perturbation theory.} 
    \textbf{(A)} Spin-valley susceptibility for the IVC$_{\rm{x}}$ and IVC$_{\rm{z}}$ phases in the absence of SOC, as a function of carrier density at $\Delta_1=17\,\rm{meV}$. 
    \textbf{(B)} 
   Stability of the IVC$_{\rm{x}}$ and IVC$_{\rm{z}}$ phases within the IVC region `2' in panel A. Between IVC$_{\rm{x}}$ and IVC$_{\rm{z}}$, the state with larger $\chi_{\tau_zs_z}$ (panel B) has lower energy in the presence of weak but non-zero SOC. 
    }
\label{fig:HFSOC}  
\end{figure}

\clearpage
\newpage
\pagebreak

\onecolumngrid

\begin{center}
\textbf{\large Supplementary information }\\[5pt]
\end{center}

\setcounter{equation}{0}
\setcounter{figure}{0}
\setcounter{table}{0}
\setcounter{page}{1}
\setcounter{section}{0}
\makeatletter
\renewcommand{\theequation}{S\arabic{equation}}
\renewcommand{\thefigure}{S\arabic{figure}}
\renewcommand{\thepage}{\arabic{page}}

\subsection{Determination of $\lambda$}\label{sec:spin-orbit-fit}

To elucidate the effects of spin-orbit coupling on the in-plane spin polarization of a VI phase we consider a toy model given by 
 \begin{equation}
     H = \frac{1}{2}\lambda \tau_z s_z  - \frac{1}{2}E_{z,\parallel} s_x  
 \end{equation}
where  $\tau$ and $s$ are the valley and spin Pauli matrices, respectively,  $E_{z,\parallel}=g\mu_B B_\parallel$, and we take $g =2$ for the spin gyromagnetic ratio. 
 The resulting in-plane magnetization, $m_\parallel$, of a valley polarized phase where $\langle\tau_z\rangle=1$ is given by 
 \begin{equation}
     m_\parallel=|m_{spin}| \frac{E_z}{\sqrt{E_z^2+\lambda^2}}
 \end{equation}

In our experiment, we measure the magnetic field evolution of the density, $n_e^*$, and displacement field, $D^*$, at which we observe first order phase transitions. The condition for a first order phase transition between two phases with contrasting in-plane magnetic moments is $E_1(n_e^*,D^*) + m_{1,\parallel} B_\parallel = E_2(n_e^*,D^*) + m_{2,\parallel} B_\parallel$. 
Experiments are performed by sweeping both $n$ and $D$ along a trajectory parameterized by $t-t_0=n+D/a$, where $t_0$ and $a$ are constants.  It follows that, at a given $B_\parallel$, the magnetic moment difference 

\begin{equation}
    \Delta m_\parallel = \frac{\Delta \mu_{1,2} + a \, \partial_D (E_1 - E_2)}{\Big[\frac{\partial B_\parallel}{\partial n^\ast} + a\,\frac{\partial B_\parallel}{\partial D^\ast} \Big]}
\label{eqn:thermodynamic_m_expanded}
\end{equation}

For the three phase boundaries studied in Fig. \ref{fig:ext_data_electron}A, C, and D, $a=0$.  For the transition studied in Fig. \ref{fig:ext_data_electron}B, we make the simplifying assumption that the two phases have similar polarizability $\partial E/\partial D$, and that $\partial B_\parallel/\partial D^*$ is negligible. As described in the main text, $\Delta \mu$ and $n_e^*(B_\parallel)$ may both be measured independently.  We find that $\Delta\mu$ is independent of $B_\parallel$ to within our experimental resolution for all four transitions, and so we may take it as a constant in our analysis. It follows that 

\begin{align}
\frac{dn_e^*}{dB_\parallel}&=\frac{\Delta m_\parallel}{\Delta \mu}\\
\Delta n_e^*(B_\parallel)&=\frac{1}{\Delta \mu}\int_0^{B_\parallel} \Delta m_\parallel dB_\parallel
    \end{align}

To fit the phase boundaries between VI and SI or IVC phases (Figs. \ref{fig:ext_data_electron}A,B, and D), we  assume $m_\parallel=m_{spin}$ for IVC and SI phases.  Then
\begin{align}
\Delta n_e^*&=\pm\frac{|m_{spin}|/(g\mu_B)}{\Delta\mu}\int_0^{B_\parallel}\left(1-\frac{E_z}{\sqrt{E_Z^2+\lambda^2}}\right) dE_z\\
\Delta n_e^*&=\pm \frac{|m_{spin}|/(g\mu_B)}{\Delta\mu}\left(E_Z+\lambda-\sqrt{E_Z^2+\lambda^2}
\right)
\end{align}

To fit the phase boundaries between VI and Sym phase (Fig. \ref{fig:ext_data_electron}B), we  assume $m_\parallel=0$ for the Sym  phase.  Then

\begin{align}
\Delta n_e^*&=\pm\frac{|m_{spin}|/(g\mu_B)}{\Delta\mu}\int_0^{B_\parallel}\left(-\frac{E_z}{\sqrt{E_Z^2+\lambda^2}}\right) dE_z\\
\Delta n_e^*&=\pm\frac{|m_{spin}|/(g\mu_B)}{\Delta\mu}\left(\lambda-\sqrt{E_Z^2+\lambda^2}
\right)
\end{align}

\subsection{Band structure and interactions}

For the band structure of rhombohedral trilayer graphene, we use the
six band model of Ref.~\cite{zhang_band_2010}:
\begin{equation}
H_{0}(\bm{k},\xi)=\left(\begin{matrix}\Delta_{1}+\Delta_{2}+\delta & \frac{1}{2}\gamma_{2} & v_{0}\pi^{*} & v_{4}\pi^{*} & v_{3}\pi & 0\\
\frac{1}{2}\gamma_{2} & \Delta_{2}-\Delta_{1}+\delta & 0 & v_{3}\pi^{*} & v_{4}\pi & v_{0}\pi\\
v_{0}\pi & 0 & \Delta_{1}+\Delta_{2} & \gamma_{1} & v_{4}\pi^{*} & 0\\
v_{4}\pi & v_{3}\pi & \gamma_{1} & -2\Delta_{2} & v_{0}\pi^{*} & v_{4}\pi^{*}\\
v_{3}\pi^{*} & v_{4}\pi^{*} & v_{4}\pi & v_{0}\pi & -2\Delta_{2} & \gamma_{1}\\
0 & v_{0}\pi^{*} & 0 & v_{4}\pi & \gamma_{1} & \Delta_{2}-\Delta_{1}
\end{matrix}\right),\label{Ham6}
\end{equation}

where $\pi=\xi k_{x}+ik_{y}$ ($\xi=\pm$ corresponds to valleys $K$
and $K'$) and the Hamiltonian is written in the basis $(A_{1},B_{3},B_{1},A_{2},B_{2},A_{3})$,
where $A_{i}$ and $B_{i}$ label the two sublattices at layer $i$.
The velocities $v_{i}$ ($i=0,3,4$) are related to the microscopic
hopping parameters $\gamma_{i}$ by $v_{i}=\sqrt{3}a_{0}\gamma_{i}/2$,
where $a_{0}=2.46$~\AA
is the lattice constant of monolayer graphene.
$\Delta_{1}$ is a potential difference between outer layers, which
is approximately proportional to the applied displacement field, while
$\Delta_{2}$ is the potential difference between the middle layer
and the average potential of the outer layers. Finally, $\delta$
is an on-site potential on $A_{1}$ and $B_{3}$. We have used the
following parameters: $\gamma_{0}=3.1\,{\rm eV},$$\gamma_{1}=0.38\,{\rm eV},$
$\gamma_{2}=-0.015\,{\rm eV}$, $\gamma_{3}=-0.29\,{\rm eV}$, $\gamma_{4}=-0.141\,{\rm eV}$,
$\delta=-0.0105\,{\rm eV}$, $\Delta_{2}=-0.0023\,{\rm eV}$. See Ref. \cite{zhou_half-_2021}
for the procedure used to fix these parameters. 

The total Hamiltonian used in our theoretical analysis is
\begin{equation}
\mathcal{H}=\sum_{\bm{k},\xi=K,K',s=\uparrow,\downarrow}\psi_{\bm{k},\xi,s}^{\dagger}H_{0}(\bm{k},\xi)\psi_{\bm{k},\xi,s}+\mathcal{H}_{{\rm C}}+\mathcal{H}_{{\rm Hunds}}+\mathcal{H}_{{\rm SOC}},
\label{eq:HFtotalHam}
\end{equation}
where $\psi_{\bm{k},\xi,s}$ is a six component spinor of annihilation
operators $\psi_{\bm{k},\xi,s,\sigma_{i}}$ that annihilate electrons
with momentum $\bm{k}$, valley index $\xi$, spin $s$, and sublattice
index $\sigma_{i}=A_{i},B_{i}$ on layer $i$. $H_{{\rm C}}$ is the
long-ranged Coulomb interaction,
\[
\mathcal{H}_{{\rm C}}=\frac{1}{A}\sum_{\bm{q}}V_{\bm{q}}
:\hat{n}_{\bm{q}}\hat{n}_{-\bm{q}}:
\]
where $\hat{n}_{\bm{q}}=
\sum_{\bm{k},\xi,s}
\psi^\dagger_{\bm{k+q},\xi,s}
\psi_{\bm{k},\xi,s}$. For the screened Coulomb interaction, we use $V_{\bm{q}}=2\pi e^2\tanh (q d)/(\epsilon q)$. Here $d=40\rm{nm}$ is the distance to the gates, and for the dielectric constant we employ $\epsilon=24$, which incorporates both the dielectric constant of the BN substrate and the effects of screening in the graphene trilayer~\cite{ghazaryan_unconventional_2021,chatterjee_inter-valley_2022,huang_spin_2023,koh_correlated_2023}.
The spatially short-ranged Hund's coupling is given by
\begin{equation}
\mathcal{H}_{{\rm Hunds}}=\frac{J_H}{A}\sum_{\bm{k},\bm{q}}
:\bm{\hat s}_{\bm{q},+}\cdot\bm{\hat s}_{-\bm{q},-}:,    
\end{equation}
where the spin density operator in valley $\xi=\pm 1$ is given by
$\bm{\hat s}_{\bm{q},\xi}
=\sum_{\bm{k},s_1,s_2}\psi^\dagger_{\bm{k+q},\xi,s_1}
\bm{s}_{s_1,s_2}\psi_{\bm{k},\xi,s_2}$ ($\bm{s}_{s_1,s_2}$ are Pauli matrices spin space).
As mentioned in the main text, the intrinsic spin-orbit coupling $\mathcal{H}_{{\rm SOC}}=\sum_{\bm{k},\xi,s}\psi_{\bm{k},\xi,s}^{\dagger}H_{\rm KM}\psi_{\bm{k},\xi,s}$ is of the Kane-Mele type, $H_{{\rm KM}}=\lambda\sigma_z\tau_zs_z/2$, where the Pauli matrices denote sublattice ($\sigma_z$), spin ($s_z$) and valley ($\tau_z$).

\subsection{Hartree-Fock analysis} 
We seek mean-field solutions for the interacting Hamiltonian Eq.~\eqref{eq:HFtotalHam} which lift the degeneracy between the four flavors spanned by the spin and valley degrees of freedom within one given band, either on the electron or hole side. To this end, we project the interactions onto a single band nearest to the Fermi level, followed by a self-consistent determination of a momentum-dependent density matrix elements in the $4\times 4$ flavor subspace~\cite{chatterjee_inter-valley_2022}.
Upon projection, the form factors of the band appear in the interaction Hamiltonian.

The Hartree-Fock calculations are performed using a rectangular momentum grid with size $61^2$, with spacing $0.003\cdot 2\pi/a$, at zero temperature. The self-consistency loop is implemented using an adaptive-step algorithm following Ref. \cite{kudin_black-box_2002}, which uses for each iteration step the linear interpolation between the previous solution and the updated solution and determines the point with minimal energy.

For spin and valley polarized states, different solutions can be enforced by selectively disabling certain matrix elements of the density matrix. In contrast, intervalley coherent states always compete with polarized solutions, and are thus more fragile numerically. The typical energy difference between a flavor symmetric state and flavor polarized ones is of the order of $1\,{\rm meV}$ per electron, while the IVC state can additionally gain around $0.05\,{\rm meV}$ over the polarized solutions. Lastly, within the IVC phase, the energy differences between the in-plane and out-of-plane spin polarizations (IVC$_\mathrm{x}$ vs. IVC$_\mathrm{z}$) are of order $1\,{\rm \mu eV}$ for $\lambda=40\,{\rm \mu eV}$, which is close to the energy resolution of our calculation (determined by how well converged the Hartree-Fock solution is). For this reason, we additionally calculate the spin-valley polarization susceptibilities for both IVC$_\mathrm{x}$ and IVC$_\mathrm{z}$, defined as $\chi_{\tau^z s^z} \langle \tau_z s_z\rangle/\lambda$. For small $\lambda$, the energy gain from SOC in the two phases is then $\Delta E_{a=X,Z} = -\frac{1}{2}\chi_{\tau^z s^z} (\lambda/2)^2$, and the phase whose susceptibility is larger is favored. We found that this method is more accurate than calculating the energy directly.  
An additional benefit of this method is that it is not necessary to specify a value for $\lambda$, as long as $\lambda$ is sufficiently small. We find that both the free energy and the spin-valley susceptibility lead to the same overall trends in the resulting phase diagrams, with only small quantitative differences.

For the determination of the phase diagram, it is necessary to first fix the value of $J_H$. By testing a few values of the Hund's coupling starting from $J_H=0$, up to $J_H=-0.02 V_{\bm{q}=0}$ while $\lambda$ is kept zero, we find that already a small $J_H=-0.011 V_0$ significantly favors the formation of IVC states. Using this value for the Hund's coupling, one can then determine the spin-valley susceptibility inside the IVC region of the phase diagram. 

The numerical results of this calculation for the electron and the hole side are shown in the main text Fig.~\ref{fig:4}B and extended data Fig.~\ref{fig:newHF} respectively. The regions of stability of the IVC$_{\rm{x}}$ and IVC$_{\rm{z}}$ within the IVC quarter metal phase are shown in Fig. \ref{fig:HFSOC}. Also shown is the spin-valley susceptibilities of the two phases along a cut through the phase diagram, showing the large enhancement of the susceptibility in the IVC$_{\rm{Z}}$ state when the phase boundary to the VI is approached. 
In the following section we show, using a phenomenological ansatz for the free energy, how the various trends in these phase diagrams can be understood.

\subsection{Model for IVC$_\mathrm{x}$ to IVC$_\mathrm{z}$ transition in quarter metal}
To understand the sequence of transitions that occurs between the IVC and the VI states within the quarter metal phase, it is useful to consider a simplified model that includes the interplay between exchange, Hund's rule, and spin-orbit interactions. 
Within the Hartree-Fock approximation, each occupied $\bm{k}$ orbital is polarized in a certain ``direction'' in spin and valley space. 
In the simplified model, we assume that this isospin polarization is nearly $\bm{k}-$independent.
We may then write the system's energy as a function of a single state $\vert \Psi\rangle$ in spin and valley space,  
\begin{equation}
\vert\Psi\rangle=\alpha\vert K\uparrow\rangle+
\beta \vert K\downarrow\rangle+
\gamma \vert K'\uparrow\rangle+
\delta \vert K'\downarrow\rangle,
\end{equation}
parameterized by the four-component spinor $\psi^{T}=(\alpha,\beta,\gamma,\delta)$. 

In general, the energy density $E$ in the quarter metal can be expressed as a function of $\psi$. We use the following phenomenological model for $E[\psi]$:
\begin{equation}
E[\psi]=\Delta_{\rm{V}}\langle\tau^{z}\rangle^{2}
+\Delta'_{\rm{V}}\langle\tau^{z}\rangle^{4}
-J_{H}\langle\bm{s}_{+}\rangle\cdot\langle\bm{s}_{-}\rangle
-\frac{\lambda}{2}\langle s^{z}\tau^{z}\rangle
-\frac{g\mu_B \bm{B}}{2}\cdot\langle\bm{s}\rangle
-\frac{g_{{\rm v}}B_{\perp}}{2}\langle\tau^{z}\rangle.  
\label{eq:Emodel}
\end{equation}
Here $\langle O \rangle \equiv \psi^\dagger O \psi$, where $O$ is a $4\times 4$ matrix in spin- and valley space.  
$\bm{\tau}$ and $\bm{s}$ are Pauli matrices in valley and spin space, respectively, and $\bm{s}_{\pm}=\frac{1\pm\tau^{z}}{2}\bm{s}$ are the spin operators projected to the two valleys. 

The parameters $\Delta_{\rm{V}}$, $\Delta'_{\rm{V}}$ represent the 
anisotropy of the energy in valley space, proportional to $n_{e}$. These parameters come from the kinetic energy and the long-ranged part of the Coulomb interactions. They are symmetric under $\rm{U_+(2)}\times \rm{U_-(2)}$ transformations corresponding to the approximate conservation of charge and spin in each valley separately. $J_H$ is the 
Hund's coupling between the spins in the two valleys 
(assumed to be ferromagnetic, $J_H\ge 0$). $\lambda$ is the strength of the Ising spin-orbit coupling. 
$J_H$ and $\lambda$ reduce the symmetry down to $\rm{U_c}(1)\times \rm{U_v}(1)\times \rm{U}_{s^z}(1)$ (corresponding to the conservation of the charge, valley, and $z$ component of the total spin). 
$\bm{B}$ is the magnetic field measured in energy units. The field couples both to the spin via the Zeeman effect, and to the perpendicular valley orbital moment (proportional to $\langle \tau^z \rangle$) with a valley $g-$factor, $g_{\rm{v}}$.

We first set $\lambda=0$, $\bm{B}=0$, and consider the phase diagram as a function of $\Delta_{\rm{V}}$ for fixed $\Delta'_{\rm{V}}$ and $J_H$. For $\Delta'_{\rm{V}}<0$, we find a first-order phase transition from the fully valley and spin polarized (VI) phase to an IVC phase at a critical value of $\Delta_{\rm{V}}$ given by $\Delta_{\rm{V},c} = -\Delta'_{\rm{V}} - J_H/4$. The IVC is stable for $\Delta_{\rm{V}}>\Delta_{\rm{V},c}$. For $\Delta'_{\rm{V}}>0$ the transition is of second order. In order to describe the experiment, where the IVC to VI transition is observed to be first order, we focus on $\Delta'_{\rm{V}}\le 0$ here. 

Next, we consider the effect of $\lambda$, assumed for simplicity to be the smallest energy scale in the problem. In the VI phase, $\lambda$ locks the spin to be perpendicular to the plane, gaining an energy $-|\lambda|$. In the IVC regime, the spin can be oriented either in the plane (which we dub the IVC$_\mathrm{x}$ phase) or out of the plane (the IVC$_\mathrm{z}$ phase). The IVC$_\mathrm{x}$ phase gains an energy $\Delta E_x = -(\lambda/2)^2/(2 J_H)$ by canting the spins slightly out of the plane. The IVC$_\mathrm{z}$ phase gains an energy by slightly imbalancing the valleys, $\langle \tau^z \rangle \ne 0$. For sufficiently small $\lambda$ such that this valley imbalance is small, the energy gain is given by $\Delta E_z = -\frac{1}{2}\chi_v (\lambda/2)^2$, where 
\begin{equation}
    \chi_v = \frac{1}{2(|\Delta'_{\rm{V}}|+\Delta_{\rm{V}} - \Delta_{1,c})}
\end{equation} is the valley susceptibility in the IVC regime (defined as $\chi_v = \langle \tau^z \rangle/h_v$, 
where $h_v$ is a small valley Zeeman field that couples to $\tau^z$, and $\langle \tau^z \rangle$ is the resulting valley polarization). 

Comparing the energies of the IVC$_\mathrm{x}$ and IVC$_\mathrm{z}$ phases, we find that the IVC$_\mathrm{z}$ is stable in a region $0<\Delta_{\rm{V}}-\Delta_{\rm{V},c}<-\left|\Delta'_{\rm{V}}\right|+\frac{J_{H}}{2}$ near the IVC to VI transition, assuming that $J_H>2|\Delta'_{\rm{V}}|$ (if $J_H<2|\Delta'_{\rm{V}}|$, IVC$_\mathrm{x}$ is the ground state for all $\Delta_{\rm{V}}>\Delta_{1,c}$). Physically, the intermediate IVC$_\mathrm{z}$ is stable if the IVC to VI transition is sufficiently weakly first order (i.e., $|\Delta'_{\rm{V}}|$ is sufficiently small), such that $\chi_v$ becomes sufficiently large upon approaching the transition from the IVC side.

\begin{figure}[ht]
    \centering    \includegraphics[width=1.0\columnwidth]{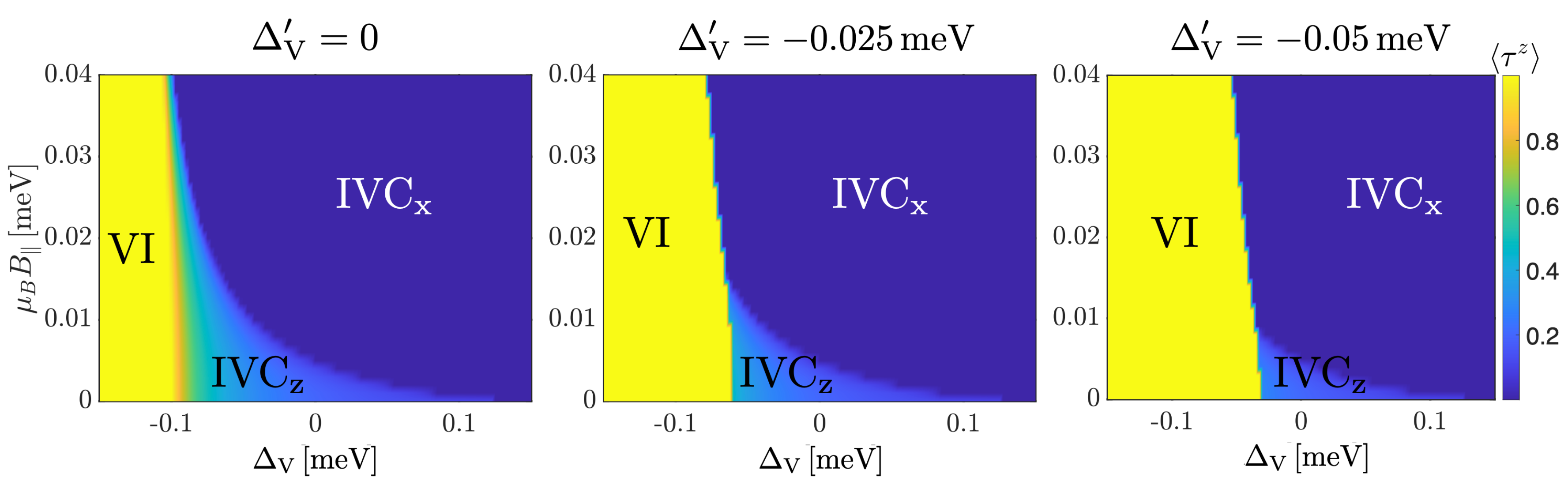}
    \caption{\textbf{Phase diagram of the theoretical model as a function of $\Delta_{\rm{V}}$ and $B_\parallel$ for different values of $\Delta'_{\rm{V}}$.} The color shows $\langle \tau_z \rangle$. In the VI phase, $|\langle \tau^z\rangle|=1$. In the IVC$_\mathrm{z}$ phase, $\langle \tau^+\rangle \ne 0$, $0<|\langle \tau^z\rangle |<1$ and $|\langle s^z\rangle|\ne 0$. In the IVC$_\mathrm{x}$ phase, $\langle \tau^+\rangle$ and $\langle s^+\rangle$ are non-zero, whereas $\langle \tau^z\rangle = \langle s^z\rangle=0$.     
    \figurelabel{fig:ext\_model}
    }
    \label{fig:ext_model}
\end{figure}

Fig. \ref{fig:ext_model} shows the phase diagram of the model in the plane $(\Delta_{\rm{V}}, B_\parallel)$ for
$\lambda = 0.2$~meV,
$J_H = 0.5$~meV,  
and different values of $\Delta'_{\rm{V}}$, obtained by minimizing the energy \eqref{eq:Emodel} over $\psi$. The color indicates $\langle \tau^z\rangle$. For $\Delta'_{\rm{V}}=0$, all the transitions are continuous, whereas for $\Delta'_{\rm{V}}<0$, the transitions from the VI to either IVC$_\mathrm{x}$ or IVC$_\mathrm{z}$ is first order. The transition from IVC$_\mathrm{x}$ to IVC$_\mathrm{z}$ is always continuous. Note how the intermediate IVC$_\mathrm{z}$ phase expands as $|\Delta'_{\rm{V}}|$ decreases; this is due to the enhxwancement of valley susceptibility $\chi_v$ in the IVC side of the IVC$_\mathrm{z}$ transition, which favors spin polarization along $z$. 
Figure \ref{fig:3}H in the main text shows phase diagrams as a function of both in-plane and out-of-plane field and $\Delta_{\rm{V}}$ calculated from this model, for $\lambda=0.2$~meV, $\Delta'_{\rm{V}} = -0.025$~meV, $g_v=5g$, $J_H=0.5$~meV. 

\subsection{Competition between IVC and valley imbalanced states}
\begin{figure}[ht]
    \centering    \includegraphics[width=0.55\columnwidth]{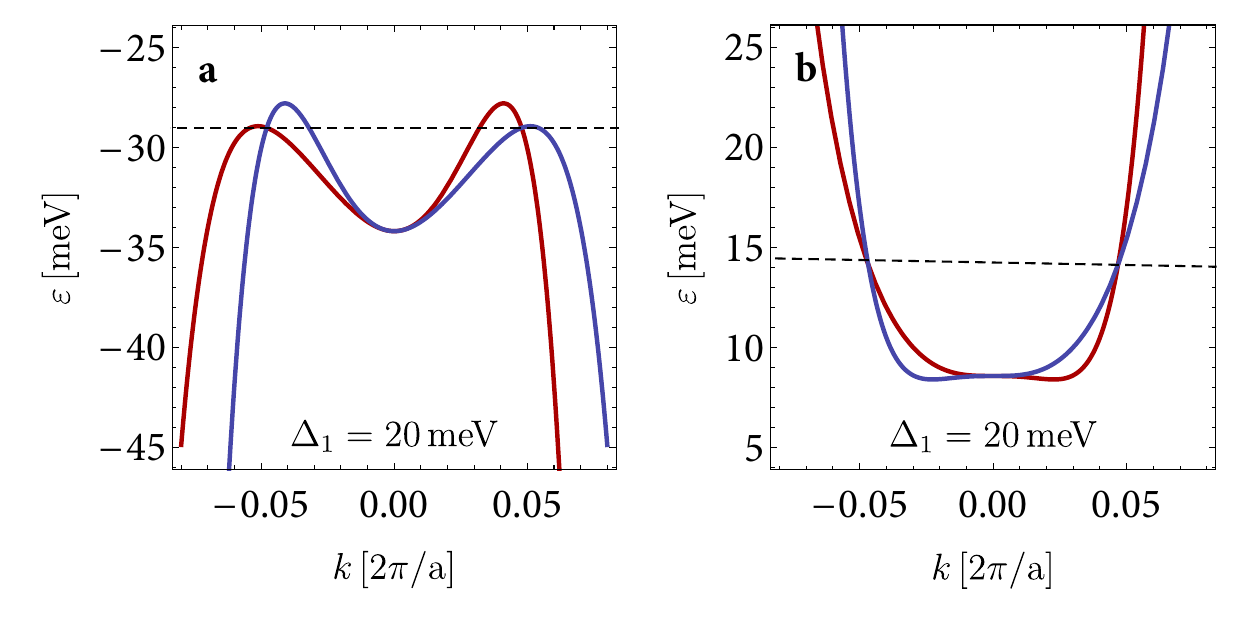}
    \caption{
    \textbf{Comparison of the non-interacting band structures of hole and electron bands.} The single-particle dispersions of valleys $K$ (blue) and $K'$ (red) for both
    \textbf{(A)} hole and \textbf{(B)} electron side, for $k_y=0$. Indicated by a dashed gray line is the crossing region, which leads to the formation of an IVC phase if the Fermi hlevel resides in the vicinity.
    \figurelabel{fig:ext\_gullys}
    }
    \label{fig:ext_gullys}
\end{figure}

Both in the experimental data and in the Hartree-Fock calculations, the order of phases as a function of increasing carrier density is dissimilar between hole doping and electron doping. Namely, on the hole side the symmetric phase at the lowest densities is followed first by an IVC, which in turn gives way to a valley imbalanced phase. On the electron side, the symmetric phase is instead first replaced by the valley imbalanced phase, and only then an IVC sets in. As long as the Hund's coupling is not too large, these differences can be understood based on the different dispersions in the electron and hole sides, as illustrated in 
Fig.~\ref{fig:ext_gullys}. 
The IVC phase is favored by the difference in kinetic energy between the two valleys at a given momentum (after folding the two valleys on top of each other), which allows for a kinetic energy gain by creating a momentum-dependent valley polarization. The VI phase is favored by the exchange energy, which is maximized for electrons in the same valley, due to the different form factors of the two valleys.

The dispersion near the bottom of the conduction band is significantly flatter than the dispersion near the top of the valence band (Fig.~\ref{fig:ext_gullys}). In particular, the crossing between the dispersions of the two valleys, where the IVC gap opens, happens closer in energy to the band edge on the hole side compared to the electron side. These differences can explain why on the hole side, the IVC phase quarter metal is favored at low hole density and the VI phase at higher hole density (where the Berry curvature enclosed by the Fermi surface is larger), whereas on the electron side the VI phase is favored at lower density (where the bands are very flat), and the IVC phase is favored at higher density (where the bands are more dispersive). 

\end{document}